\documentclass[aps,prd,reprint,superscriptaddress,titlepage,a4paper,showkeys]{revtex4-2}
\usepackage{graphicx}
\usepackage{amsmath}
\usepackage{amsfonts}
\usepackage{amssymb}
\usepackage{xcolor}
\usepackage{xspace}
\usepackage[colorlinks=true, allcolors=blue]{hyperref}
\usepackage[capitalise]{cleveref}
\usepackage{threeparttable}
\usepackage{multirow}
\usepackage{tikz-feynman}
\usepackage{enumerate}
\usepackage{subfig}
\usepackage{textgreek}
\usepackage{float}
\usepackage{soul}

\newcommand{\mmu}{{\text{M}}}

\newcommand{\ammu}{{\overline{\text{M}}}}

\begin{document}
\title{High-Precision Physics Experiments at Huizhou Large-Scale Scientific Facilities}

\author{FengPeng~An}
\affiliation{School of Physics, Sun Yat-sen University, Guangzhou 510275, China}

\author{Dong~Bai}
\affiliation{College of Mechanics and Engineering Science, Hohai University, Nanjing 211100, China}

\author{Hanjie~Cai}
\affiliation{State Key Laboratory of Heavy Ion Science and Technology, Institute of Modern Physics, Chinese Academy of Sciences, Lanzhou 730000, China}
\affiliation{University of Chinese Academy of Sciences, Beijing 100049, China}

\author{Siyuan~Chen}
\affiliation{School of Physics, Sun Yat-sen University, Guangzhou 510275, China}
\affiliation{Platform for Muon Science and Technology, Sun Yat-sen University, Guangzhou 510275, China}

\author{Xurong~Chen}
\affiliation{State Key Laboratory of Heavy Ion Science and Technology, Institute of Modern Physics, Chinese Academy of Sciences, Lanzhou 730000, China}
\affiliation{School of Nuclear Science and Technology, University of Chinese Academy of Sciences, Beijing 100049, China}
\affiliation{Southern Center for Nuclear Science Theory, Institute of Modern Physics, Chinese Academy of Sciences, Huizhou 516000, China}

\author{Hongyue~Duyang}
\affiliation{Shandong University, Jinan, China, and Key Laboratory of Particle Physics and Particle Irradiation of Ministry of Education, Shandong University, Qingdao 266237, China}


\author{Leyun~Gao}
\affiliation{State Key Laboratory of Nuclear Physics and Technology, School of Physics, Peking University, Beijing, 100871, China}

\author{Shao-Feng~Ge}
\affiliation{State Key Laboratory of Dark Matter Physics, Tsung-Dao Lee Institute \& School of Physics and Astronomy, Shanghai Jiao Tong University, Shanghai 200240, China}
\affiliation{Key Laboratory for Particle Astrophysics and Cosmology (MOE) \& Shanghai Key Laboratory for Particle Physics and Cosmology, Shanghai Jiao Tong University, Shanghai 200240, China}

\author{Jun~He}
\affiliation{School of Physics and Technology, Nanjing Normal University, Nanjing 210097, China}

\author{Junting~Huang}
\affiliation{School of Physics and Astronomy, Shanghai Jiao Tong University}

\author{Zhongkui~Huang}
\affiliation{State Key Laboratory of Heavy Ion Science and Technology, Institute of Modern Physics, Chinese Academy of Sciences, Lanzhou 730000, China}
\affiliation{University of Chinese Academy of Sciences, Beijing 100049, China}

\author{Igor~Ivanov}
  \affiliation{School of Physics and Astronomy, Sun Yat-sen University, Zhuhai 519082, P.R. China}   

\author{Chen~Ji}
\affiliation{Key Laboratory of Quark and Lepton Physics (MOE) and Institute of Particle Physics, Central China Normal University, Wuhan 430079, China}
\affiliation{Southern Center for Nuclear Science Theory, Institute of Modern Physics, Chinese Academy of Sciences, Huizhou 516000, China}

\author{Huan~Jia}
\affiliation{State Key Laboratory of Heavy Ion Science and Technology, Institute of Modern Physics, Chinese Academy of Sciences, Lanzhou 730000, China}
\affiliation{University of Chinese Academy of Sciences, Beijing 100049, China}

\author{Junjie~Jiang}
\affiliation{School of Physics and Astronomy, Shanghai Jiao Tong University}


\author{Xiaolin Kang}
\affiliation{School of Mathematics and Physics, China University of Geosciences, Wuhan 430074, China}
\author{Soo-Bong~Kim}
\affiliation{School of Physics, Sun Yat-sen University, Guangzhou 510275, China}

\author{Chui-Fan~Kong}
\affiliation{State Key Laboratory of Dark Matter Physics, Tsung-Dao Lee Institute \& School of Physics and Astronomy, Shanghai Jiao Tong University, Shanghai 200240, China}
\affiliation{Key Laboratory for Particle Astrophysics and Cosmology (MOE) \& Shanghai Key Laboratory for Particle Physics and Cosmology, Shanghai Jiao Tong University, Shanghai 200240, China}

\author{Wei~Kou}
\affiliation{State Key Laboratory of Heavy Ion Science and Technology, Institute of Modern Physics, Chinese Academy of Sciences, Lanzhou 730000, China}
\affiliation{School of Nuclear Science and Technology, University of Chinese Academy of Sciences, Beijing 100049, China}

\author{Qiang~Li}
\affiliation{State Key Laboratory of Nuclear Physics and Technology, School of Physics, Peking University, Beijing, 100871, China}

\author{Qite~Li}
\affiliation{State Key Laboratory of Nuclear Physics and Technology, School of Physics, Peking University, Beijing, 100871, China}

\author{Jiajun~Liao}
\affiliation{School of Physics, Sun Yat-sen University, Guangzhou 510275, China}

\author{Jiajie~Ling}
\affiliation{School of Physics, Sun Yat-sen University, Guangzhou 510275, China}

\author{Cheng-en~Liu}
\affiliation{State Key Laboratory of Nuclear Physics and Technology, School of Physics, Peking University, Beijing, 100871, China}

\author{Xinwen~Ma}
\affiliation{State Key Laboratory of Heavy Ion Science and Technology, Institute of Modern Physics, Chinese Academy of Sciences, Lanzhou 730000, China}
\affiliation{University of Chinese Academy of Sciences, Beijing 100049, China}

\author{Hao~Qiu}
\affiliation{State Key Laboratory of Heavy Ion Science and Technology, Institute of Modern Physics, Chinese Academy of Sciences, Lanzhou 730000, China}



\author{Jian~Tang}
\affiliation{School of Physics, Sun Yat-sen University, Guangzhou 510275, China}
\affiliation{Platform for Muon Science and Technology, Sun Yat-sen University, Guangzhou 510275, China}



\author{Rong~Wang}
\affiliation{State Key Laboratory of Heavy Ion Science and Technology, Institute of Modern Physics, Chinese Academy of Sciences, Lanzhou 730000, China}
\affiliation{School of Nuclear Science and Technology, University of Chinese Academy of Sciences, Beijing 100049, China}

\author{Weiqiang~Wen}
\affiliation{State Key Laboratory of Heavy Ion Science and Technology, Institute of Modern Physics, Chinese Academy of Sciences, Lanzhou 730000, China}
\affiliation{University of Chinese Academy of Sciences, Beijing 100049, China}

\author{Jia-Jun~Wu}
\affiliation{School of Physical Sciences, University of Chinese Academy of Sciences, Beijing 100049, China}
\affiliation{Southern Center for Nuclear Science Theory, Institute of Modern Physics, Chinese Academy of Sciences, Huizhou 516000, China}

\author{Jun~Xiao}
\affiliation{Shanghai EBIT Laboratory, Key Laboratory of Nuclear Physics and Ion-Beam Application (MOE), Institute of Modern Physics, Fudan University, Shanghai 200433, PR China}

\author{Xiang~Xiao}
\affiliation{School of Physics, Sun Yat-sen University, Guangzhou 510275, China}

\author{Yu~Xu}
\affiliation{Advanced Energy Science and Technology Guangdong Laboratory, Huizhou 516000, China}
\affiliation{Institute of Modern Physics, Chinese Academy of Sciences, Lanzhou 730000, China}

\author{Weihua~Yang}
\affiliation{College of Nuclear Equipment and Nuclear Engineering, Yantai University, Yantai, Shandong 264005, China}

\author{Xiaofei~Yang}
\affiliation{State Key Laboratory of Nuclear Physics and Technology, School of Physics, Peking University, Beijing, 100871, China}

\author{Jiangming~Yao}
\affiliation{School of Physics and Astronomy, Sun Yat-sen University, Zhuhai 519082, P.R. China}   

\author{Ye~Yuan}
\affiliation{Institute of High Energy Physics, Chinese Academy of Sciences,Beijing 100049, China}
\affiliation{School of Physical Sciences, University of Chinese Academy of Sciences, Beijing 100049, China}

\author{Mushtaq~Zaiba}
\affiliation{State Key Laboratory of Heavy Ion Science and Technology, Institute of Modern Physics, Chinese Academy of Sciences, Lanzhou 730000, China}
\affiliation{School of Nuclear Science and Technology, University of Chinese Academy of Sciences, Beijing 100049, China}

\author{Pengming~Zhang}
\affiliation{School of Physics and Astronomy, Sun Yat-sen University, Zhuhai 519082, P.R. China}  

\author{Shaofeng~Zhang}
\affiliation{State Key Laboratory of Heavy Ion Science and Technology, Institute of Modern Physics, Chinese Academy of Sciences, Lanzhou 730000, China}
\affiliation{University of Chinese Academy of Sciences, Beijing 100049, China}

\author{Shuo~Zhang}
\affiliation{Advanced Energy Science and Technology Guangdong Laboratory, Huizhou 516000, China}

\author{Shihan~Zhao}
\affiliation{School of Physics, Sun Yat-sen University, Guangzhou 510275, China}
\affiliation{Platform for Muon Science and Technology, Sun Yat-sen University, Guangzhou 510275, China}

\author{Liping~Zou}
\affiliation{Sino-French Institute of Nuclear Engineering and Technology, Sun Yat-sen University, Zhuhai 519082, China}

\date{\today}


\clearpage

\begin{abstract}

In response to the capabilities presented by the High-Intensity Heavy Ion Accelerator Facility (HIAF) and the Accelerator-Driven Subcritical System (CiADS), as well as the proposed Chinese Advanced Nuclear Physics Research Facility (CNUF), we are assembling a consortium of experts in relevant disciplines—both domestically and internationally—to delineate high-precision physics experiments that leverage the state-of-the-art research environment afforded by CNUF. Our focus encompasses six primary domains of inquiry: hadron physics—including endeavors such as the super eta factory and investigations into light hadron structures; muon physics; neutrino physics; neutron physics; the testing of fundamental symmetries; and the exploration of quantum effects within nuclear physics, along with the utilization of vortex accelerators. We aim to foster a well-rounded portfolio of large, medium, and small-scale projects, thus unlocking new scientific avenues and optimizing the potential of the Huizhou large scientific facility. The aspiration for international leadership in scientific research will be a guiding principle in our strategic planning. This initiative will serve as a foundational reference for the Institute of Modern Physics in its strategic planning and goal-setting, ensuring alignment with its developmental objectives while striving to secure a competitive edge in technological advancement. Our ambition is to engage in substantive research within these realms of high-precision physics, to pursue groundbreaking discoveries, and to stimulate progress in China's nuclear physics landscape, positioning Huizhou as a preeminent global hub for advanced nuclear physics research.

\end{abstract}

\maketitle
\clearpage


\section{Introduction}

Large ion accelerator facilities play a pivotal role as essential tools in the domain of nuclear physics.  The Institute of Modern Physics at the Chinese Academy of Sciences is embarking on the construction of two pivotal infrastructures: the High-Intensity Heavy Ion Accelerator Facility (HIAF) and the Accelerator-Driven Subcritical System (CiADS), both anticipated to commence operations from 2025 to 2027. Under the purview of the 14th Five-Year Plan and the strategic roadmap for the 2035 National Major Scientific Infrastructure Development Plan, the Institute is further proposing the ``Chinese Advanced Nuclear Physics Research Facility" (CNUF). This endeavor aims to leverage the capabilities of HIAF and CiADS through a phased development strategy during the 14th (2026-2030) and 15th (2031-2035) Five-Year Plans, positioning CNUF as a premier multidisciplinary facility in nuclear physics research.

\begin{table}[]
    \centering
    \caption{Typical beam parameters for the HIAF-BRing}
    \begin{tabular}{lcc}
    \hline
    \hline
    Ions & Energy (GeV/u) & Beam Intensity (ppp) \\
    \hline
    $^{238}\rm{U}^{35+}$ & 	0.835 & $2.0 \times 10^{11}$ \\
    $^{209}\rm{Bi}^{31+}$ & 	0.85 & $2.4 \times 10^{11}$ \\
    $^{78}\rm{Kr}^{19+}$ & 	1.7 & $6.0 \times 10^{11}$ \\
    $^{12}\rm{C}^{6+}$ & 	2.6 & $1.2 \times 10^{12}$ \\
    proton & 	9.3 & $6.0 \times 10^{12}$ \\
    \hline
    \hline
    \end{tabular}
    \label{tab:BRing}
\end{table}

\begin{table}[]
    \centering
    \caption{Proton beam parameters of CiADS Linac}
    \begin{tabular}{lcccc}
    \hline
    \hline
    Parameters & Design & 2028 & 2030 & Upgrade ability\\
    \hline
    Energy (MeV) & 500 & 600 & 600 & 1500 \\
    Current (mA) & 5 & 0.5 & 5 & 10 \\
    Power (MW)   & 2.5 & 0.3 & 3 & 15 \\
    \hline
    \hline
    \end{tabular}
    \label{tab:CiADS}
\end{table}

In this paper, we outline a targeted experimental agenda designed to harness the world-class research environments provided by HIAF, CiADS, and the upcoming CNUF. Our focus encompasses six principal categories of high-precision experiments: investigations into new physics beyond the Standard Model via rare decays of $\eta$ mesons; studies in muon and neutrino physics; examinations of cold neutron interactions; tests of fundamental symmetries; explorations of quantum effects in nuclear physics; and advances in vortex ion physics. Our objective is to establish a dynamic balance among large, medium, and small-scale projects to open new scientific avenues, thereby maximizing the impact of the Huizhou infrastructure.

\section{Huizhou Hadron Spectrometer} 

HIAF can provide full ion beam species ranging from protons to $^{238}\rm{U}$, with
 proton beam energy up to 9.3 GeV or $^{238}\rm{U}^{76+}$ beams up to 2.45 GeV/u. 
For its upgrade to CNUF, the facility can accelerate the proton beam to an energy of 25 GeV or $^{238}\rm{U}^{76+}$ beam to 7.3 GeV/u.
This provides good opportunities for a wide range of physical research, including the search for new particles and new interactions beyond the standard model, the test of basic symmetries, the search for new (exotic) hadron states, exotic di-baryons and new (multi-strange) hypernuclei, precise measurement of hadron and hypernucleus prop-
erties, the search and location of the nuclear matter phase boundary and critical point, etc. 
In order to conduct these physics studies, we plan to build an experimental spectrometer at the HIAF high energy terminal, called the Huizhou Hadron Spetrometer (HHaS).
If acquiring sufficient funding support, we plan to take about 3 years to develop the key detector technologies for HHaS and another about 5 years to construct the whole spectrometer. 
Successful construction of HHaS will promote the development of medium- and high-energy particle physics and nuclear physics researches based on facilities in China.

\subsection{Conceptual Design}

\begin{figure}[t]
    \centering
    \includegraphics[width=\columnwidth]{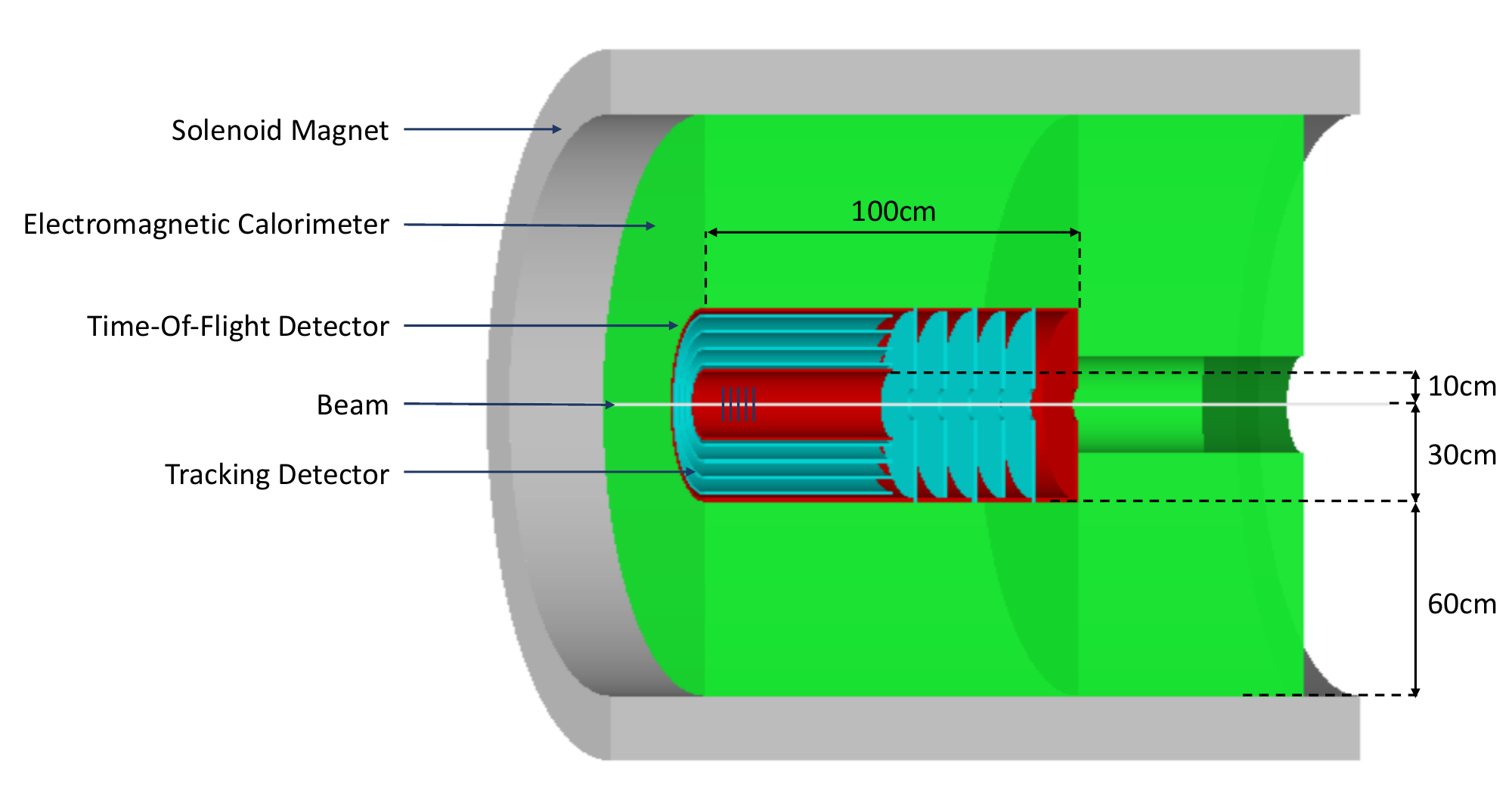}
    \caption{\label{fig:HHaS}HHaS conceptual design.}
\end{figure}

The conceptual design of HHaS is shown in Fig.~\ref{fig:HHaS}. 
The solenoidal magnet forms the backbone of the spectrometer, enabling precise momentum measurement of charged particles. The magnetic field can be adjusted between 0.8 and 1.6 Tesla according to physics requirements. 

The 5-dimensional tracking system features an all-silicon pixel detector with $\sim$100 μm pitch pixels, incorporating both energy deposition and time measurement capabilities on each pixel.
With $\sim<$20-μm hit position resolution and a 1.6-Tesla magnetic field, $\sim$3\% momentum resolution can be achieved for particle $p/m>$1 with a tracker radius of merely 30 cm. This makes HHaS a very compact and cost-effective spectrometer.
The good hit resolution also provides exceptional vertex resolution, which is crucial for clear reconstruction of strange particles and hypernuclei.
The 10-ns level time resolution enables separation of hits from different collisions when HHaS runs at an ultra-high event rate of around 100 MHz.
The measurement of energy deposition helps distinguishing different nuclei with the same charge over mass ratio, which is crucial for hypernucleus and nuclear matter studies.

In and out of the pixel tracking detector, the Low-Gain Avalanche Detectors (LGAD) are placed for time-of-flight measurements. A remarkable time resolution of about 30 ps can be achieved with state-of-the-art LGAD technology, which helps the identification of various charged particles and further separation of tracks from different collisions. 

The dual-readout electromagnetic calorimeter is placed outside the time-of-flight detector. Layers of lead glass Cherenkov detectors and plastic scintillators are assembled alternatively to read out Cherenkov light and scintillation light at the same time, providing excellent separation between electromagnetic and hadronic showers. This technology, adopted from the REDTOP collaboration~\cite{REDTOP:2022slw}, can achieve an energy resolution of approximately 3\% at 1 GeV and can distinguish clearly photons vs. neutrons and electrons vs. pions. 

HHaS accepts particles with transverse momenta above 50 MeV/c from 10 to 100 degrees in polar angle with full azimuthal coverage.
It can clearly identify and measure $\pi^{\pm}$, $K^{\pm}$, $p$, $\bar{p}$, d, $^3$He, $^4$He, $^6$Li, $e^{\pm}$ and $\gamma$.
Most impressively, the system can record interaction events with rates of $\sim$100 MHz for proton beam experiments and $\sim$1 MHz for heavy ion experiments, a capability that sets it apart from many existing experimental set-ups. 
For example, the typical event rates for ATLAS, CMS, ALICE, STAR and HADES experiments are 75 kHz, 100 kHz, 50 kHz, 1 kHz and 20 kHz, respectively. 
These comprehensive abilities make HHaS a versatile spectrometer capable of many physics studies to be discussed in the following subsections.

\subsection{Search for Beyond-Standard-Model Physics and Test of Fundamental Symmetries with $\eta$ Meson Decays}

The search for Beyond-Standard-Model (BSM) physics is one of the most exciting frontiers of particle physics. While the Standard Model (SM) has been extremely successful in explaining the fundamental particles and interactions observed so far, there remain unanswered questions, particularly concerning dark matter, neutrino masses, and violations of fundamental symmetries~\cite{Arbey:2021gdg,Bertone:2018krk,Aramaki:2015pii}.

The $\eta$ meson, a neutral pseudoscalar meson, is an excellent candidate for probing BSM physics particularly through its deacy channels. The physics potential of $\eta$ mesons in BSM searches has been explored extensively, especially in the context of dark matter, neutrino masses, and discrete symmetry violations. Theoretical studies, including those from the Bonn/Julich group~\cite{Mai:2021vsw, Ronchen:2015vfa}, predict that $\eta$ meson decays can be sensitive to new particles and interactions not predicted by the SM, such as dark photons, axion like particles, and dark scalars~\cite{ParticleDataGroup:2020ssz}. The decay modes are particularly valuable for searching for dark matter candidates and testing CP violations and lepton flavor symmetries.

Recent experimental efforts, such as the search for dark photons at BESIII~\cite{Ding:2024iqr} and the study of axion-like particles at LHC~\cite{Alonso-Alvarez:2023wni}, demonstrate the potential for finding new physics in meson decays. The BESIII collaboration, for example, found significant results from rare decays of $J/\psi \ \text{and} \ \phi \ \text{mesons}$ that set stringent limits on the dark photon parameters. These studies emphasize the importance of precise measurements in high-luminosity facilities.
The recent work of the REDTOP collaboration~\cite{REDTOP:2022slw} highlights the importance of $\eta$ rare decays, such as $\eta \rightarrow \gamma \gamma^* \rightarrow e^+ e^- \gamma$, $\eta \rightarrow S\pi^0 \rightarrow \pi^+ \pi^- \pi^0$, and $\eta \rightarrow S\pi^0 \rightarrow e^+ e^- \pi^0$, in revealing dark photons and dark scalar particles. 
Recent theoretical calculations also suggest that measurable signals of dark photons and dark scalars could be detected in $\eta$ meson decays at the energy levels available at HIAF~\cite{Lanfranchi:2020crw}.

The combination of advanced spectrometer, high luminosity beams for large $\eta$ meson event collection, low background conditions, and the ability to accurately measure rare decays makes HHaS@HIAF an ideal facility for probing dark matter candidates.
This is justified by simulation studies which shows that with HHaS at HIAF, more than one order of magnitude higher sensitivities can be achieved for dark photon and dark scaler searches within certain mass ranges, and the CP violation test is two orders of magnitude more precise than similar existing results~\cite{Chen:2024wad}.
With the completion of the HHaS, these studies are set to open new avenues of research, contributing to the global effort to uncover physics beyond the SM.

\subsection{Light Hadron Physics}
Understanding how quarks combine into hadrons remains a central challenge in modern physics. While quantum chromodynamics (QCD) governs the strong interaction, its nonperturbative nature at low energies makes it difficult to tackle, leaving the strong force the least understood among the four fundamental interactions. The hadron spectrum serves as a crucial probe, driving major experimental efforts at BES III, Belle II, LHCb and J-PARC experiments. High-precision measurements are essential for progress, and advanced facilities like HIAF and the upcoming CNUF are expected to make significant contributions. HHaS@HIAF’s broad energy coverage makes it an ideal platform for light hadron studies, allowing comprehensive investigations of hadronic interactions.

These facilities provide high-intensity proton beams, potentially including polarized beams, which enable efficient production and investigation of light baryon resonances. 
This capability offers a robust platform for investigating the ``missing resonance" problem. While nucleon resonances have been widely studied, multi-strange hyperons remain a greater challenge. Despite their discovery in the 1950s-60s, the Particle Data Group (PDG) confirms only a fraction of predicted states: of 23 $\Lambda$ states, only 14 are established; $\Sigma$ and $\Xi$ have 9 and 11, respectively; and $\Omega$ has just three. Moreover, key properties such as spin-parity assignments remain undetermined for many $\Xi$ and $\Omega$ states, even though lattice QCD and quark models predict around 70 states in this region~\cite{ParticleDataGroup:2024cfk}. 
Usually, the yield of $\Xi^*$ and $\Omega^*$ resonance states is extremely low.
The high-intensity potentially-polarized proton beam at HIAF and the unprecedented high event rate of HHaS enable effective production and detection of these states.
In addition, the potential secondary kaon beam at HIAF also provides an efficient mechanism to produce hyperons, which is similar to the J-PARC secondary kaon beam experiments.
Therefore, the high-intensity proton beam on HIAF and the secondary kaon beam that may be produced in the future bring significant advantages to the study of these difficult-to-detect states.

HHaS@HIAF also offers an
excellent opportunity to study light exotic hadrons and probe anomalies such as
the mass flip of $N$(1535) and Roper resonance. This anomaly is
believed to arise from significant five-quark admixtures in its structure,
challenging the conventional three-quark picture~\cite{Liu:2005pm}. 
Furthermore, the internal structure of $\Lambda(1405)$ remains a hotly debated
topic~\cite{ParticleDataGroup:2024cfk}. Competing models suggest it might be either a conventional three-quark
state or a molecular-like state formed by strong $\bar{K}N$ interactions.
Precise experimental data—especially on its line shape and decay patterns—are
essential to resolve this issue, and HHaS@HIAF is well equipped to meet this
challenge.
In addition, the search for exotic states such as the dibaryon $d^*(2380)$
continues to be a major objective. Although initial hints have emerged from
experiments like WASA@COSY, independent confirmation is still lacking~\cite{WASA-at-COSY:2011bjg}. With its
high-intensity proton beams, HHaS@HIAF holds great potential for delivering
definitive insights into these elusive states and for uncovering further
dibaryon resonances, deepening our understanding of the strong interaction in
the nonperturbative regime.

\subsection{Nuclear Matter at High Baryon Density}
Heavy-ion collision experiments at RHIC and LHC can produce nuclear matter reaching temperatures of several trillion degrees, creating and studying Quark-Gluon Plasma (QGP) in laboratory conditions - the hottest matter ever produced by humankind. QGP existed in the early universe, making its study profoundly significant for cosmology. After approximately two decades of research, we have gained substantial understanding of QGP properties produced at high collision energies.

When collision energies are reduced, the system's baryon density and baryon chemical potential increase. Theoretical predictions suggest this may lead to a first-order phase transition and a critical point~\cite{Fu:2019hdw}. This represents an important research direction in heavy-ion collision physics, motivating major international efforts including RHIC's beam energy scan program, and the construction of new facilities like NICA and FAIR. Heavy-ion collisions at HIAF also provide excellent opportunities to discover and locate the predicted phase transition and critical point.

Near the critical point, theoretical models predict that critical fluctuations would cause oscillatory or divergent behavior in higher-order moments of conserved charges~\cite{Stephanov:2011zz}. However, STAR collaboration's measurements of net-proton higher-order moments at energies above 7.7 GeV have not observed the predicted oscillatory behavior~\cite{thestarcollaboration2025precisionmeasurementnetprotonnumber}. This suggests the critical point may exist at lower energies and higher baryon densities, which is also consistent with several recent theoretical predictions~\cite{Hippert:2023bel}.

Consequently, the energy ranges covered by HIAF and CNUF ($E_{k} <$ 9.1 GeV/u, $\sqrt{s_{NN}} <$ 4.5 GeV) represent a crucial region for searching the phase transition critical point.
With its MHz-level event rate capability for heavy-ion collisions, large acceptance and good particle identification ability, HHaS offers unique advantages for measuring critical fluctuation observables - including higher-order moments of conserved charges and light nucleus yield ratios - that are predicted to be sensitive to the critical point.

\subsection{Hypernucleus Physics}
Hypernucleus contains hyperons in addition to the usual protons and neutrons. 
These exotic systems allow study of hyperon-nucleon interactions that cannot be probed in normal nuclei, providing key insights into the strong interaction and strange quark behavior in nuclear environments, which is crucial for the understanding neutron star interiors where hyperons may exist.

Heavy-ion collisions at several GeV energy can create a lot of hyperons, which may coalesce with nucleons in the collision system and form hypernuclei.
STAR@RHIC, ALICE@LHC and future CBM@FAIR uses or plans to use heavy-ion collisions to study light hypernuclei. 
Using heavy-ion beams with energies up to 9.1 GeV/u at HIAF, light hypernuclei can be created abundantly and studied.
The good hit resolution of HHaS provides exceptional vertex resolution, which is crucial for eliminating almost all combinatorial backgrounds when reconstructing hypernuclei.
HHaS' 5-dimensional pixel tracking detector can also measure the energy loss of charged particles, which can be used to distinguish different nuclei with the same charge over mass ratio, such as d, $^4$He and $^6$Li. This is important for the identification of the daughter particles of hypernuclei.
Taking advantage of the high luminosity of HIAF and the unprecedented event rate of HHaS, various properties of hypernuclei, such as lifetime, mass, decay branching fractions, as well as their production yield, transverse momentum spectra and azimuthal anisotropy can be measured precisely, and new (multi-strange) hypernuclei can also be searched.

\section{Muon Physics}
Muon ($\mu^-$), as well as its antiparticle ($\mu^+$), has spin $1/2$ and a rest mass of 105.658~MeV/$c^2$. The lifetime of muon is about $2.2~\mu s$, making it the lightest unstable charged lepton. Taking this advantage, muon is regard as a promising probe for the new physics BSM. Moreover, it has broad applications in various areas of material science~\cite{Ning:2025hai}. However, there is no operating accelerator muon source in China at present. The proposed muon beamlines based on the large scientific facilities under construction, HIAF and CiADS, are expected to break through the bottleneck~\cite{Xu:2025spd,Cai:2023caf}. Some representative topics can be brought up in the context of the forthcoming muon beamlines in China.

\begin{figure*}[t]
    \centering
    \subfloat[\label{mace-phasei}MACE Phase-I]{\includegraphics[width=0.3\textwidth]{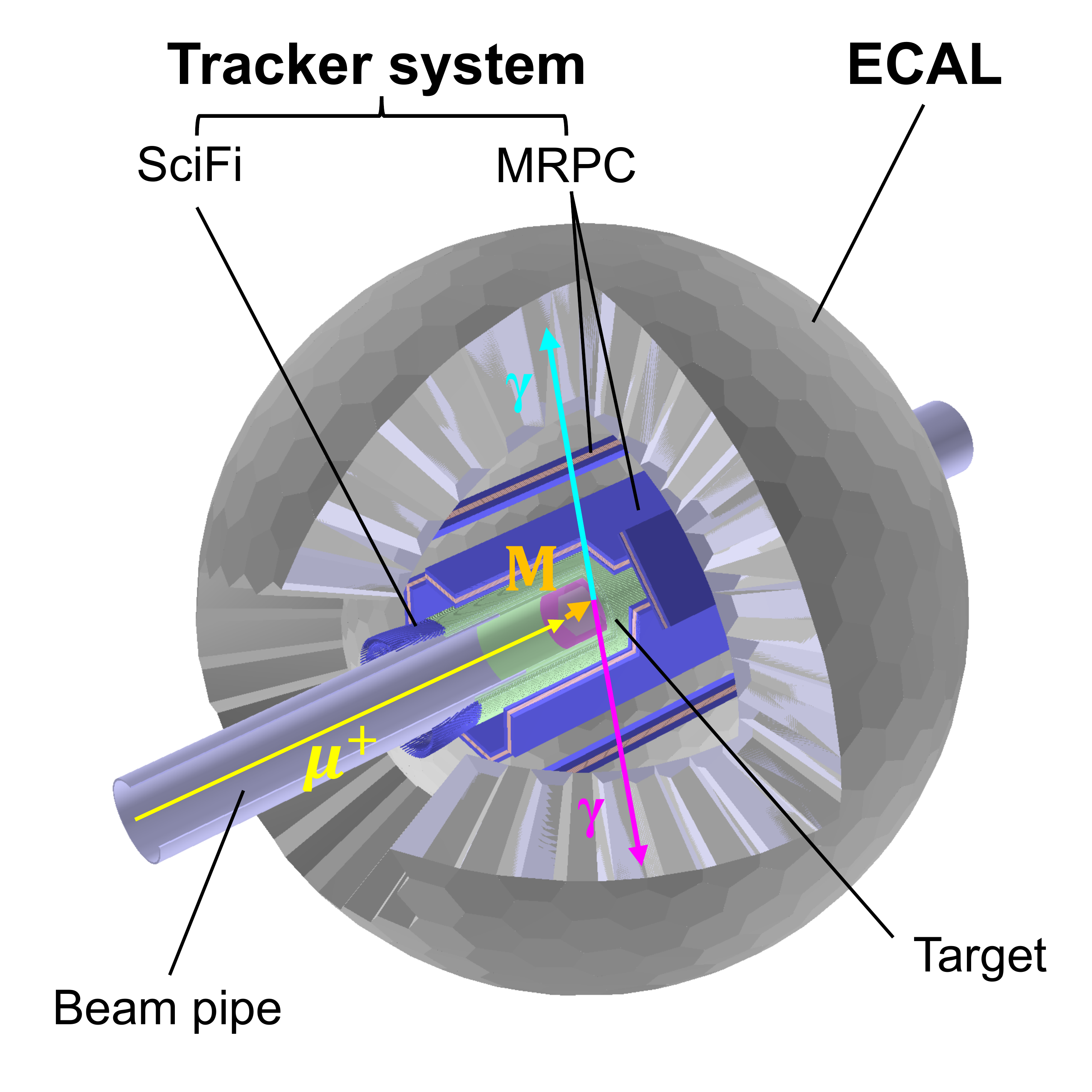}}
    \hfil
    \subfloat[\label{mace-phaseii}MACE Phase-II]{\includegraphics[width=0.65\textwidth]{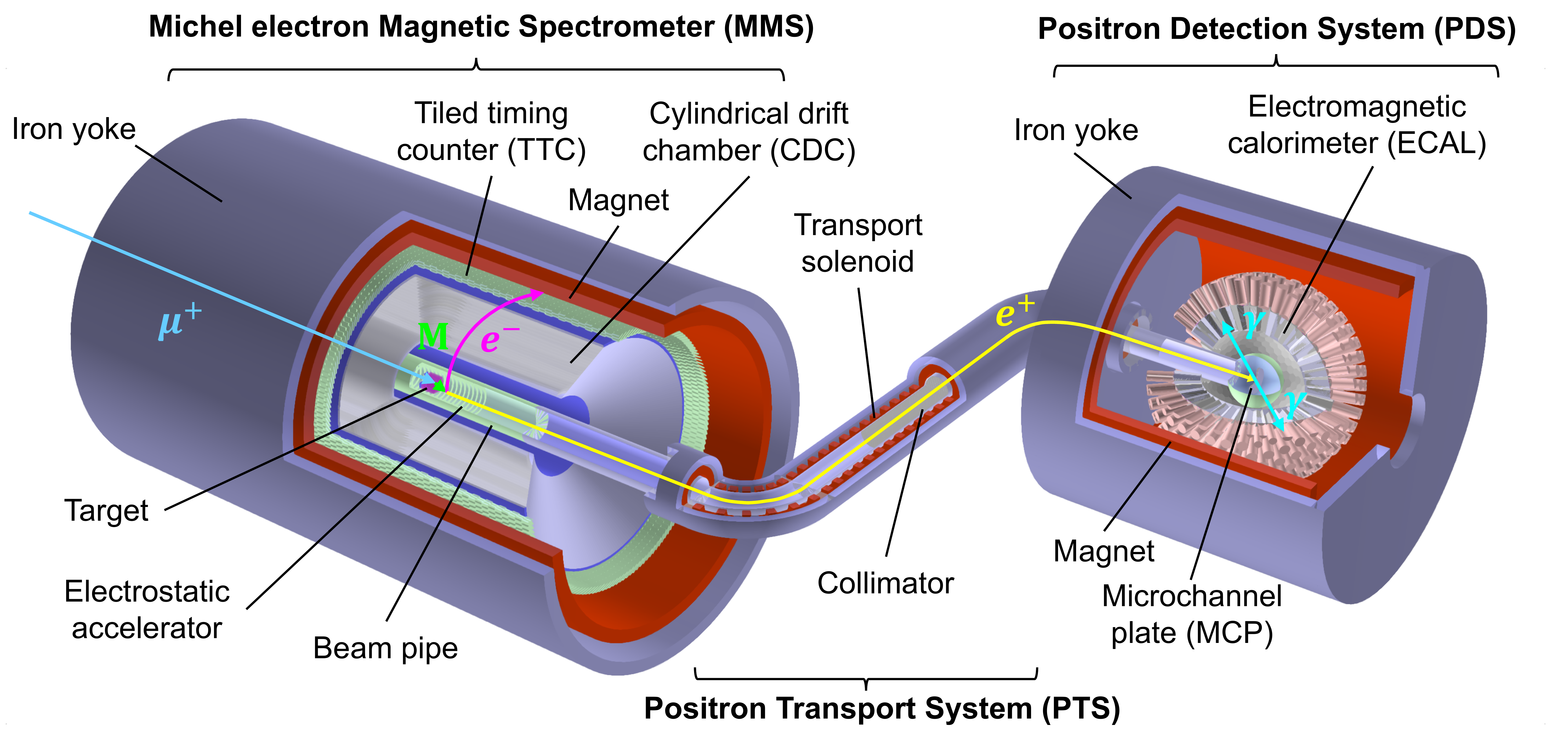}} \caption{\label{fig:mace-detector-concept}MACE detector concept.}
\end{figure*}

\subsection{Charged lepton flavor violation}
In the SM, lepton flavor is a strictly conserved quantum number. The discovery of neutrino oscillations, which confirms that neutrinos have mass, provides the first direct evidence for physics BSM, demonstrating that {neutral leptons could exhibit flavor mixing effects}. Consequently, there is strong motivation to pursue experimental searches for BSM physics via charged lepton flavor violation (cLFV) processes.
Popular channels include $\mu^+ \rightarrow e^+\gamma$, $\mu^+ \rightarrow e^+e^-e^+$, and $\mu^-N \rightarrow e^-N$~\cite{Perrevoort2023}. In recent years, notable experiments searching for these cLFV channels include MEG and its upgrade, MEG-II~\cite{Baldini2018}, Mu3e~\cite{Arndt2021} at Paul Scherrer Institute, COMET in Japan~\cite{Abramishvili2020,Xu:2024btf}, and Mu2e in the USA~\cite{Bartoszek2014}. Due to the lack of a dedicated muon beamline, no muon cLFV experiments have been proposed in China thus far. However, with the ongoing construction of the Huizhou large scientific facility, the research and development of muon sources is now underway. It paves the way for exploring unique muon cLFV processes in the coming years.

\paragraph{Muonium-to-antimuonium conversion.}
Muonium (also identified as $\mmu$) is a bound state consists of a positively charged muon and an electron. In 1960, it was firstly discovered by Vernon~W.~Hughes~\textit{et al.}~\cite{PhysRevLett.5.63}. Unlike the mentioned processes above which only violate the lepton flavor by one unit, the muonium-to-antimuonium conversion is a $\Delta L_\ell=2$ process, providing important complement rather than the $\Delta L_\ell=1$ cLFV processes. 
Theoretical analyses of the conversion probability have been performed, both in particular new physics models~\cite{Pontecorvo:1957cp,Feinberg:1961zza,ClarkLove:2004,CDKK:2005,Li:2019xvv,Endo:2020mev,Han:2021nod,Fukuyama:2021iyw}, and the Standard Model effective field theory (SMEFT)~\cite{Conlin:2020veq}. Observation of muonium converting into anti-muonium will provide a ``smoking gun'' evidence of new physics in the leptonic sector~\cite{Bernstein:2013hba,Willmann:1998gd}.

In 1999, the MACS experiment at PSI reported the most recent upper limit on the conversion probability, which is $P \lesssim 8.3 \times 10^{-11}$ at 90\% confidence level. However, this result has remained unchallenged by any experiment over the past two decades. The Muonium-to-Antimuonium Conversion Experiment (MACE) is proposed and expected to discover this cLFV process or further constrain its conversion probability~\cite{Bai:2024skk}.

\cref{fig:mace-detector-concept} shows the concept of the MACE detector apparatus. If the conversion occurs, the antimuonium will soon decay into a Michel electron and a low-energy positron. To identify the signal of $\text{M}$-to-$\overline{\text{M}}$ conversion, the designed detector system includes three primary components: a Michel electron magnetic spectrometer (MMS), a momentum-selection positron transport system (PTS), and a positron detection system (PDS). The Michel electron magnetic spectrometer consists of a cylindrical drift chamber (CDC), a set of tiled timing counters (TTC), and a magnet. It is designed to measure the tracks and momenta of high-energy Michel electrons. The positron transport system consists of an electrostatic accelerator and a positron transport solenoid. The PTS is used to transport low-energy positrons to the positron detection system, while the transverse positions of the positrons is conserved. The positron detection system includes a microchannel plate (MCP) and an electromagnetic calorimeter (ECAL)~\cite{Chen:2024jmg}. The low-energy positron will be detected by the MCP, and then annihilate into two back-to-back gamma rays which will be detected by the ECAL.

In the baseline run plan, MACE will utilize a muon beamline with $10^8$~s$^{-1}$ muon-on-target ($\mu$OT) flux, beam momentum of 24~MeV/$c$ and an root mean square (RMS) spread of 1.35~MeV/$c$. The data acquisition duration will be set at 1 year. As a result, MACE is expected to collect $1.20 \times 10^{14}$ muonium decay events in the vacuum. With the current MACE detector design and run plan, if no signal events are observed, it is expected to achieve a conversion probability upper bound, at 90\% confidence level, of
\begin{equation}
\begin{aligned}
    P(\mmu \to \ammu)\lesssim 3.8\times 10^{-13}~,
\end{aligned}
\end{equation}
which suggests an more than 2 orders of magnitude improvement in sensitivity compared to the MACS experiment at PSI. SMEFT calculations demonstrates that MACE has the potential to probe new physics at energy scales comparable to or beyond those of $\mu$TRISTAN, a future muon collider~\cite{Hamada:2022mua,Bai:2024skk}.

\paragraph{Muon rare decays.}
In addition to the muonium-to-antimuonium conversion process, other barely touched channels in muon cLFV decays are also of interest. MACE \textsc{Phase-I} (\cref{mace-phasei}) aims at the muon neutrinoless double radiative decay $\mu^+\to e^+\gamma\gamma$ and the muonium annihilation $\mathrm{M} \to \gamma\gamma$, which are less concerned but interesting. It has been reported by several theoretical studies that the rate of the most popular cLFV channel $\mu^+\to e^+\gamma$ could suffer a stronger suppression than $\mu^+\to e^+\gamma\gamma$~\cite{PhysRev.126.375,Fortuna:2022sxt,Uesaka:2024tfn}, offering a solid motivation to search for this process. Worth mentioning, the search of an exotic boson $X$ in the $\mu^+\to e^+X$ process is also possible for the \textsc{Phase-I} detector by reconstruction the decay products of $X$, i.e. $X\to\gamma\gamma$~\cite{Heeck:2017xmg,Renga:2019mpg}.

The Crystal Box experiment at the Los Alamos Meson Physics Facility (LAMPF) was the latest to constrain the branching ratio of \(\mathcal{BR}(\mu^+ \to e^+ \gamma \gamma) < 7.2 \times 10^{-11}\) at the 90\% confidence level in 1988~\cite{Bolton:1988af}.
The latest experiment searching for the muonium annihilation $\mathrm{M} \to \gamma\gamma$ was conducted by C. M. York et al. in 1959~\cite{PhysRevLett.3.288}. The experiment merely setup a loose bound of $\mathcal{O}(10^{-5})$ on $\mu^+ e^-\to \gamma\gamma$~\cite{PhysRevLett.3.288,Hughes1975}. No updated results or dedicated experimental plans have been proposed for these channels, as the limited geometrical acceptance of the current MEG calorimeter constrains its capability to probe the aforementioned processes. With the optimized detector system in MACE \textsc{Phase-I}, sensitivities of $\mathcal{O}(10^{-12}-10^{-13})$ are expected for the two decay modes mentioned above.

\paragraph{Muon-to-electron conversion.}
The aim of current experiments COMET \& Mu2e is to reach the sensitivity of $10^{-17}$ level. After them, new experiment base on CNUF could reach $10^{-19}$ level.

Conducting a next generation muon-to-electron conversion experiment requires a pulsed beam. At CiADS, a beam with a pulsed time structure similar to the design of COMET/Mu2e can be obtained through the chopper system of the Medium Energy Beam Transport (MEBT) line of the linear accelerator. For such a pulsed beam, the extracted beam power at CiADS can reach 0.5 MW, corresponding to a number of protons on target (POT) per pulse of up to $7 \times 10^9$.  The single event sensitivity can be calculated according to the following equation
	
\begin{align}
       B (\mu^- \, + \text{Al} \rightarrow \, e^{-} \, + \text{Al}) \, = \, \frac{1}{N_{\mu}\cdot f_{\text{cap}}\cdot f_{\text{geo}}} \,,
\end{align}
where $f_{\text{cap}}$ is the muon capture efficiency at the muon stopping target, $f_{\text{geo}}$ is the geometrical acceptance of the detector. 
For the muon LFV program, current experiments face limitations primarily in two areas: statistics and systematics, respectively:
1. Experiments such as MEGII and Mu3e are primarily limited by accidental coincidence backgrounds, which are a form of systematic error. Consequently, simply increasing the beam intensity (to improve statistics) will not substantially enhance their sensitivity or improve the upper limits.
2. In contrast, experiments like COMET and Mu2e \cite{Mu2e-II:2022blh} are mainly constrained by the available muon flux, representing statistical errors. Since these setups do not suffer from significant accidental backgrounds, systematic uncertainties can be minimized through careful facility design and precise simulations. Therefore, increasing the muon beam intensity is expected to significantly improve their sensitivity.
Looking ahead, more intense muon fluxes are anticipated from upcoming accelerator facilities in China. By applying the same physics processes--such as muon-to-electron conversion used in COMET and Mu2e--extrapolations based on projected beam intensities are both ambitious and plausible. These developments hold the potential to push the boundaries of LFV searches to new, more stringent limits.

Search for new particles such as Majoron etc., can also be carried out on such Muon-to-electron conversion experiment facility and the sensitivity can reach the order $O(10^{-10})$ \cite{Majoron2023}.

\subsection{High-Precision Spectroscopy of Muonic Atoms}

\begin{figure}[t]
    \centering
    \includegraphics[width=0.95\columnwidth]{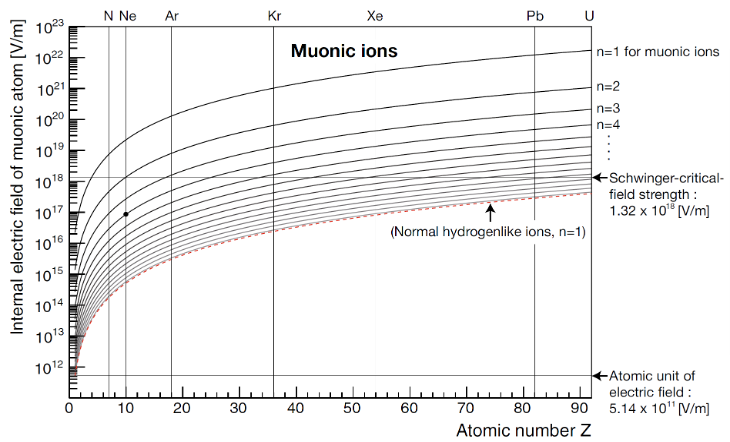}
    \caption{Internal electric field strength of muonic atom for each principal quantum numbers from n = 1 to 14, and of normal hydrogenlike atom for n = 1, as a function of atomic number Z. (Reprinted with permission from \cite{Okada:2020fjo}.) }
\end{figure}

\subsubsection{Testing QED in extreme fields using muonic atoms.} 
Quantum electrodynamics (QED), the quantum field theory that describes the interaction between light and matter, is commonly regarded as the best-tested quantum theory in modern physics. However, this claim is mostly based on extremely precise studies performed in the domain of relatively low field strengths and light atoms and ions \cite{Safronova:2017xyt}. Testing QED effects in extreme electromagnetic fields is one of the frontiers in atomic physics. Significant progresses have been achieved in experimental investigations of the Lamb shift, the hyperfine splitting, and the bound-electron g factor of highly charged ions (HCI) in the last decade \cite{PhysRevLett.94.223001,ncomms8.15484,Morgner:2023nature}. However, testing QED in these extreme fields is limited by the uncertainties due to finite nuclear size effects (FNS) and nuclear polarization or deformation of HCI \cite{PhysRevLett.86.3959,Kozlov:2018mbp}. 

Muonic atoms, where a negative muon is captured onto the atomic orbitals, replacing an electron, provide extremely strong fields because muons are 207 times closer to the nucleus than electrons and thus the internal electric field strength is to be 40,000 times higher than that of normal atoms. The electric field is proportional to the cube of the atomic number Z, as shown in Fig. 3. As a result, high-resolution X-ray measurement of muonic atoms was proposed as an ideal probe to explore QED under extremely strong electric fields \cite{Paul:2020cnx}. By precisely measuring the energy of muonic characteristic X-rays emitted when muonic atoms are deexcited from a specific level to lower levels, QED can be verified under a strong electric field. 

It is well known that the effect of vacuum polarization is particularly significant in quantum electrodynamic effects on muonic atoms. With the development of superconducting transition edge sensor (TES) microcalorimeter, a proof-of-principle experiment for testing strong-field QED has been performed by measuring the X-ray spectroscopy of muonic neon atoms at J-PARC \cite{PhysRevLett.130.173001}. A high-intensity muon source driven by a continuous-wave superconducting linac is planned at CiADS which will significantly advance muonic atom x-ray spectroscopy \cite{Cai:2023caf}. We plan to employ the latest developed TES microcalorimeter in the energy range from 10 keV to 200 keV \cite{Zhang:2022rlr}, which is developed at the institute of Modern physics, to conduct precision X-ray measurements of muonic atoms for testing the QED effect under extreme strong electromagnetic field conditions. In the X-ray muonic atom experiment, a low-density gas target will be required to avoid rapid refilling of electrons into this system from surrounding atoms, as described in \cite{Paul:2020cnx,PhysRevLett.130.173001}.

\subsubsection{Study of nuclear properties with muonic atoms.} Since 2010, a highly precise measurement of the proton charge radius using, for the first time, muonic hydrogen spectroscopy unexpectedly led to controversy, as the value disagreed with the previously accepted one, called “proton size puzzle” \cite{Pohl:2010zza}. Since then, atomic and nuclear physicists have been trying to understand this discrepancy by checking theories, questioning experimental methods and performing new experiments \cite{Xiong:2019umf,Karr:2020wgh}. 

Muonic atom spectroscopy offers unprecedented precision for measuring nuclear electromagnetic properties. When a negative muon (207 times heavier than an electron) replaces an orbital electron, its reduced Bohr radius causes significant wavefunction overlap with the nucleus. This proximity enhances nuclear structure effects by approximately seven orders of magnitude compared to electronic atoms, enabling precise determination of charge radii, magnetic moments, and higher-order electromagnetic distributions. The precision measurement of the X-rays emitted during the formation process of a muonic atom has a longstanding history in probing the shape and size of nuclei \cite{Antognini:2020mva,Saito:2022fwi}. Precision measurements of absolute charge radii provide important benchmarks and inputs for modern nuclear structure theory using realistic nuclear potentials \cite{Mizuno:2023jak}.

Significant progress has been achieved through several pioneering experimental programs.  The CREMA collaboration extracted precise charge radii of $^{1,2}$H and $^{3,4}$He through laser spectroscopy of $\mu^{1,2}$H, and $\mu^{3,4}$He$^{+}$~\cite{Pohl:2010zza,Antognini:2013,Pohl:2016sc,Krauth2021,Schuhmann:2023} and is now extending these studies to muonic lithium. The $\mu$-X experiment focuses on high-resolution muonic atom X-ray spectroscopy of radioactive or rare isotopes using microgram target materials, both extracting nuclear charge radii and searching for charge-parity violation signals~\cite{Adamczak:2022jbo,Skawran:2019krq,Knecht:2020npz}. Complementarily, the QUARTET collaboration employs metallic magnetic calorimeters (MMC) for high-precision X-ray spectroscopy of light muonic atoms, targeting charge radii determinations from Li to Ne isotopes with unprecedented accuracy~\cite{Ohayon:2023hze,Unger:2023ppv}.

In addition to the light-mass region, high-precision charge radii measurements of stable or long-lived radioactive nuclei in the middle- and heavy-mass regions are also valuable. Currently, fundamental nuclear physics research focuses heavily on the properties and structure of exotic nuclei~\cite{Ye:2024slx}. Charge radii serve as sensitive probes for studying the exotic phenomena that emerge in such nuclei. To date, most charge radii measurements of unstable nuclei have been conducted via isotope shifts using laser spectroscopy~\cite{Yang:2022wbl}. However, in most of the cases, the precision of the extracted nuclear charge radii is limited not by experimental resolution but by uncertainties in the atomic mass shift and field shift factors. For isotopic chains with more than three stable isotopes, these atomic factors can be well calibrated if the charge radii of stable nuclei are measured with high precision—for example, through muonic atom spectroscopy or electron scattering. It should be noted that the measurements of the absolute charge radii of high-Z radioactive elements from high-precision X-ray spectroscopy are complementary to the measurements of relative differences in mean-square radii along the isotopic chain available from laser spectroscopy \cite{Campbell:2016yxe,Yang:2022wbl}. 

The recently planned high-intensity muon beam at CiADS \cite{Cai:2023caf} will provide a significantly advance by measuring X-rays in higher-Z muonic atoms to investigate nuclear properties. The energy resolution of X-ray spectroscopy for muonic atoms will be improved with a factor of 5-10 by employing the advanced cryogenic TES and MMC micro-calorimeter technology as compared to conventional semiconductor detectors. Therefore, the nuclear properties, such as nuclear charge radii, magnetic dipole moments, and Zemach radii, can be investigated with a high precision across a broader range of isotopes. These measurements will provide unique experimental conditions for studying the properties of atomic nuclei and critical benchmarks for testing ab initio nuclear structure calculations. Taking full advantages of the current advanced experimental techniques and potential muonic atom capabilities at new facilities, the precision of charge radii measurements for stable nuclei across the nuclear chart could be significantly improved. This advancement would undoubtedly benefit charge radii studies of unstable nuclei as well.

Facilities at J-PARC, RIKEN-RAL, and PSI have concentrated on measuring the hyperfine splitting in $\mu$H to extract the proton Zemach radius---a parameter characterizing the convolution of proton's charge and magnetization distributions that provides critical insights into nuclear spin structure~\cite{Sato:2015gra,FAMU:2023rqk,Amaro:2021goz}. The planned high-intensity muon beam at CiADS would enable hyperfine structure determinations with relative precision of 0.1--1 ppm via munoic atom X-ray spectroscopy. Priority targets will include muonic Li and Be isotopes, which allow high-precision benchmarking with the Zemach radii determined through spectroscopy of their electronic counterparts~\cite{Qi:2020,Li:2020,Dickopf:2024rmq}. These measurements are essential for determining nuclear Zemach radii across a broader range of isotopes. CiADS also will significantly advance muonic atom spectroscopy by extending precision measurements to hyperfine splitting in higher-$Z$ muonic atoms. Implementing the advanced MMC micro-calorimeter, these measurements will provide critical benchmarks for testing ab initio nuclear structure calculations, particularly for nuclear polarizability contributions to muonic atom spectra that currently limit the precision in the determination of nuclear radii from muonic atom spectroscopy.


\subsection{Probing and Knocking with Muons(PKMu)}

Muons serve as valuable probes bridging fundamental physics and applications. Both cosmic and accelerator-based muons enable exploration beyond the Standard Model(SM), which, despite its success, cannot account for neutrino masses or dark matter(DM). The absence of any direct detection of Weakly Interacting Massive Particles (WIMPs) to date has prompted the scientific community to explore alternative models, including light-mass or muon-philic DM.~\cite{Essig:2022dfa, Harris:2022vnx, Bai_2014}.

Muon scattering remains underexplored. Early studies(1960s-1980s) targeted nuclear structure~\cite{Drees:1983pd}, while recent efforts like MEG~\cite{MEG:2016leq} and Mu3e~\cite{Hesketh:2022wgw} focus on rare muon decays. Scattering experiments, such as NA64~\cite{NA64:2024klw} and the planned MUonE~\cite{CarloniCalame:2015obs}, now reemerge to test new physics. Different beam energies probe distinct regimes, highlighting muon scattering's potential.

\begin{figure}
\centering
\includegraphics[width=1.\columnwidth]{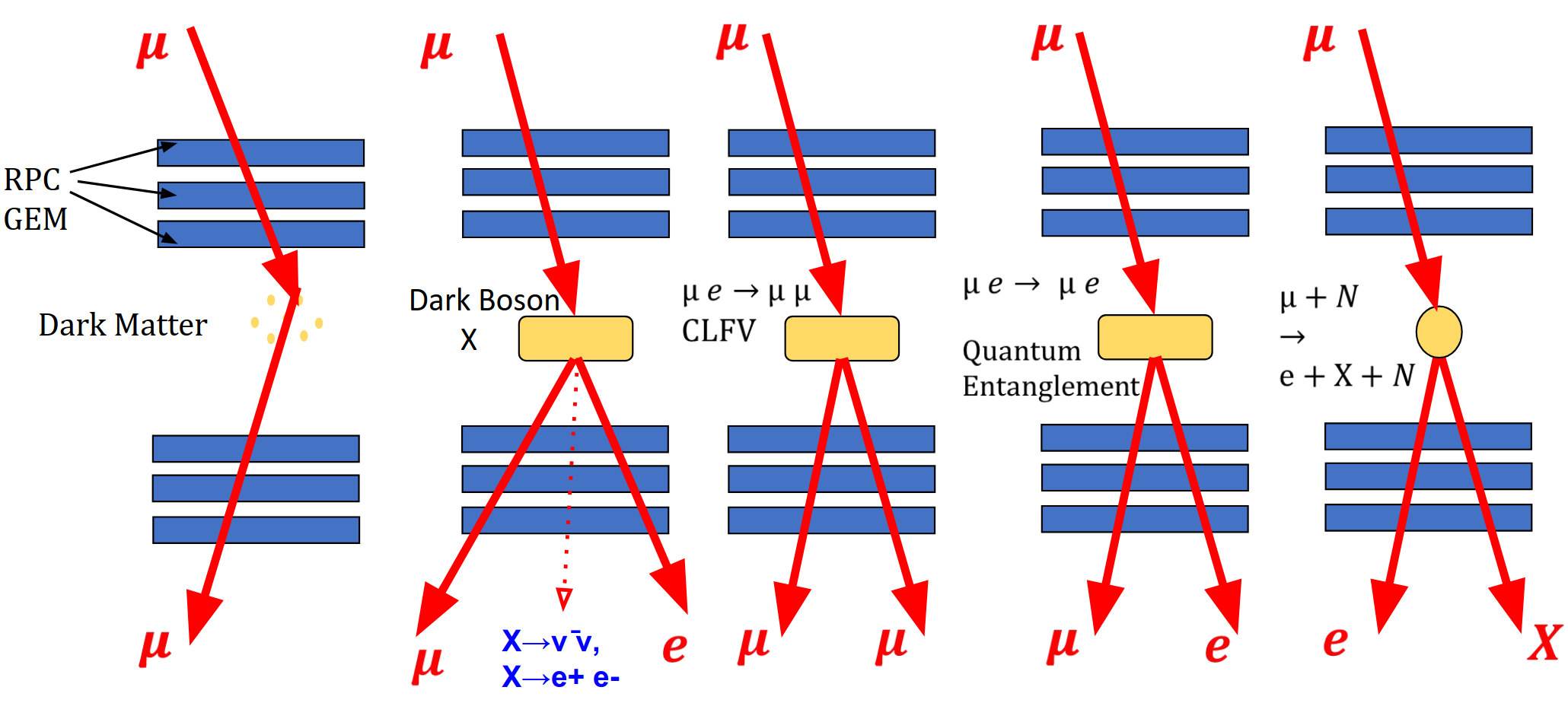}  
\caption{A schematic diagram of the PKMu project for exploring DM, dark bosons, CLFV, and quantum entanglement phenomena through probing and knocking with muons~\cite{gao2025probingknockingmuons}}
\label{fig:pkmu}
\end{figure}

The High Energy Fragment Separator beamline at HIAF will deliver GeV-scale muon beams~\cite{Xu:2025spd}. As shown in Fig.~\ref{fig:pkmu}, this enables studies on muon-philic DM~\cite{Ruzi:2023mxp, Yu:2024spj}, charged lepton flavor violation(CLFV)~\cite{Gao:2024xvf}, and quantum correlations in $\mu-e$ scattering~\cite{Gao:2024leu, Gao:2025kdi}.

PKMu experiment will employ Gas Electron Multiplier (GEM)~\cite{gemd} or Resistive Plate Chamber (RPC)~\cite{Riegler:2004gc} detectors for muon tracking. GEMs provide precise spatial resolution, widely applied in high-energy physics, including CMS~\cite{Abbas:2022fze, Pellecchia:2022lsd}. RPCs offer timing precision, robustness, and stability. Together, they offer a cost-effective platform for high-resolution tracking.

\textbf{Dark Matter}: We propose a model-independent DM search using cosmic muons tracked by GEMs and RPCs. Simulations with \textsc{Geant4}~\cite{GEANT4:2002zbu} suggest muon trajectory shifts upon DM scattering. The CRY program~\cite{CRY} provides a 3–4 GeV average sea-level muon spectrum.Using the aforementioned tools, we simulated the scattering processes between muons and both atmospheric background and dark matter particles of varying masses. The results show distinguishable differences from scattering in a detector filled only with air~\cite{Yu:2024spj}. Such differences allow extraction of dark matter cross-section limits through detailed background simulations, and can also deepen studies of secondary cosmic-ray components and their scattering behavior. Assuming a local DM density of $\rho \sim 0.3$~GeV/$\rm{cm}^3$ and a detector volume of $V \sim 1\ {\rm m}^3$, the estimated number of DM particles is $3 \times 10^{5(6,4)}$. For a one-year run, the expected sensitivity of the muon-philic dark matter scattering cross section is:
\begin{equation}\label{eq:muonbox}
\sigma_{\mu,\mathrm{DM}} \sim 10^{-12(-13,-11)}~{\rm cm}^2.
\end{equation}
This approach complements nuclear recoil searches(e.g., XENON1T, PandaX)~\cite{Yu:2024spj, PhysRevLett.131.011005}. Detection efficiency remains key to improving limits. To further enhance the sensitivity of DM searches, we can conduct experiments untilizing muon beams with higher intensity and better focusing characteristic. It is crucial to emphasize that even a null result from this search would still be significant in placing new constraints on such scenarios. It would tighten the bounds on muon-philic DM interaction strengths.

\textbf{Light Dark Bosons}: Thermal freeze-out models for sub-GeV DM require light mediators, such as a dark $Z^\prime$ gauge boson~\cite{Battaglieri:2017aum}. The $L_\mu - L_\tau$ model~\cite{Foot:1991mn, He:1991pn, He:1991qd, Foot:1994vd}, predicts $Z^\prime$ coupling weakly to muons. We target the process $\mu e^- \to \mu e^- X$($X$ decays invisibly), characterized by large-angle muon and electron scattering with minimal downstream signals. The 1–10 GeV muon beam at HIAF~\cite{Xu:2025spd}(operational$\sim2025-2026$) enables this study, with superior sensitivity to $Z^\prime$ around 10 MeV compared to CERN's 160 GeV. Detailed discussions can be found in~\cite{gao2025probingknockingmuons}.

\textbf{Quantum Entanglement}: Quantum entanglement, a hallmark of quantum theory, has been observed in photons, leptons, and top quarks~\cite{Horodecki:2009zz, Bell:1964kc, Barr:2024djo, ATLAS:2024fsd, CMS:2024pts, CMS:2024zkc}. In muon-electron and positron-electron scattering, entanglement is quantified via concurrence and CHSH inequality~\cite{Gao:2024leu}. Simulations show strong CHSH violations at 1-10 GeV muon energies, with event rates of $\sim2.6 \times 10^{4}/day$ for 10 GeV muons and $\sim1.9 \times 10^9/s$ for 1 GeV positrons. Detailed discussions can be found in~\cite{Gao:2024leu}.

In Bhabha scattering, spin correlations yield near-Bell states, observable via secondary scattering. Even with 20\% polarized targets, only a few hundred seconds of data suffice for state discrimination~\cite{Gao:2024leu}.

\section{Neutrino Physics}

\subsection{CP Violation}

The observed baryon asymmetry in our Universe is a long-standing
issue \cite{Canetti:2012zc}. Namely, there are a lot of matter
in our Universe but almost no anti-matter. Otherwise, the matter
and anti-matter should annihilate with each other. Understanding
the origin of baryon asymmetry is of profound importance for
understanding the existence of the matter world in our Universe
and the cosmic evolution. One possible explanation is the
leptogenesis mechanism \cite{Fukugita:1986hr,Buchmuller:2005eh,Davidson:2008bu} that has deep connection with
neutrinos and their mixing. The CP violation is a key element
for leptogenesis \cite{Sakharov:1967dj}. In the standard three
generation framework, the neutrino PMNS mixing matrix contains
a Dirac CP phase $\delta_{D}$ that can be measured by oscillation
experiments. Observing $\delta_D$ with oscillation experiments
can then provide a very important support for the leptogenesis
mechanism.

After the non-zeo $\theta_{13}$ was measured by Daya Bay in 2012 \cite{DayaBay:2012fng}, the neutrino physics
has entered a new era. This opens the possibility for
experimental measurement of the Dirac CP phase $\delta_D$. 
The current measurements by T2K indicates that
the leptonic CP is maximally violated wtih $\delta_D$ around
$-90^\circ$ \cite{T2K:2023smv} for both mass orderings
while NOvA prefers $180^\circ$ ($-90^\circ$) for the normal
ordering (inverted ordering) \cite{NOvA:2021nfi}. However, the
sensitivity of these two experiments is not high enough to fully
determine the value of $\delta_D$. Especially, running both
the neutrino and anti-neutrino modes at the first oscillation
peak has several issues such as splitting the beam
time, degeneracy between CP values, and intrinsically large
CP uncertainty for the maximal CP phase
\cite{Evslin:2015pya,Ge:2017qqv,Smirnov:2018ywm,Ge:2020xkm,Ge:2022iac,Ge:2024mco}.
The next-generation experiments with optimization are desirable to
establish the existence of leptonic CP violation and
provide a precise measurement of $\delta_{D}$.


As one of the most powerful accelerator facilities, HIAF could generate high-intensity
proton beams.
The flux intensity and frequency can reach 
$4\times 10^{14}$\,ppp (proton per pulse) and
3\,Hz at the upgraded HIAF (HIAF-U). The proton 
beam is expected to reach $3.78\times 10^{22}$\,POT/year under 
the assumption of a 100\% duty factor. The hit
of protons on target can first produce $\pi^\pm$ and $K^\pm$
which further decay to muons $\mu^\pm$ that can
continue the decay chain. Neutrinos are produced
from the pion and muon decays. 
A simple neutrino beamline is setup in the 
simulation, which includes a graphite target with a length 
of 70\,cm, a solenoid employed for concentrating pions, an 
8-degree dipole magnet utilized to select pions with the 
desired energy, and an 80\,m-long decay tunnel for the 
generation of neutrinos. By adjusting 
the magnetic field of the dipole magnet, it becomes feasible
to modulate the peak of the neutrino flux spectrum. 
Moreover, efforts are being made to further optimize the 
neutrino beamline, which is expected to further increase
the number of neutrinos on the JUNO target.
In order to optimize
the CP sensitivity, the peak energy of the neutrino
beam shall peak around $\sim500\,$MeV.

\begin{figure}[t]
\includegraphics[width=1\columnwidth]{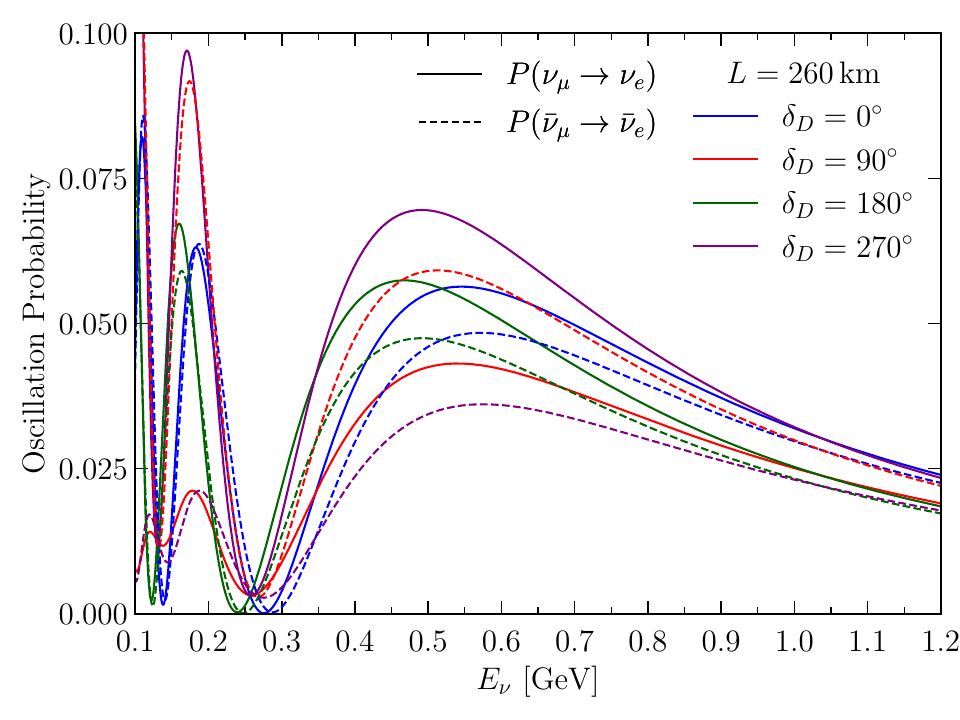}
\caption{Oscillation probability at HIAF-to-JUNO.}
\label{fig:prob-nu}
\end{figure}

The distance between the HIAF site and the Jiangmen
Underground Neutrino Observatory (JUNO) \cite{JUNO:2015zny}
is 260\,km. 
The neutrino oscillation probability during
the propagation is shown in Fig.\,\ref{fig:prob-nu}
where the peak energy of HIAF neutrinos is close to
the first oscillation peak. Since the reactor neutrinos
are below 10\,MeV while the HIAF neutrinos are mainly
above 100\,MeV, there is no overlap with each other.
Hence, the accelerator neutrino experiment from HIAF
to JUNO can run simultaneously with the original JUNO
reactor neutrino experiment. In addition, the HIAF beam
has a narrow pulse of $\sim 400\,$ns to significantly
reduce the atmospheric neutrino background to make
to make it neglibigle. Especially, the JUNO
detector is capable of identifying those high-energy
neutrinos with electron and muon flavors which is
essential for the neutrino CP measurement. The particle
identification efficiency for both flavors can reach
around 80\% while the mis-identification rate is roughly
at the percentage level. With a 20\,kt fiducial mass of the JUNO
detector, the high-intensity HIAF neutrino beam can 
produce $\mathcal{O}(100)$ $\nu_\mu(\bar\nu_\mu)\rightarrow\nu_e(\bar\nu_e)$
events per year.

If established, the HIAF-to-JUNO project would be the
first accelerator neutrino experiment in China. 
With high-intensity neutrino beams and optimized neutrino energies, HIAF-to-JUNO would be sensitive to the Dirac CP phase.
Especially for the maximal CP phases at 
$\pm90^\circ$, its uncertainty tends to be 
largest inside the whole CP range for 
other accelerator experiments such as 
T2K/T2HK/NOvA/DUNE that are not optimized for the
CP uncertainty. This leaves the HIAF-to-JUNO project
a chance of optimizing its sensitivity for
the maximal CP phase that is indicated by current
measurements. This would allow HIAF-to-JUNO to
become a very promising experiment for the precision
CP measurement.

\subsection{Coherent Scattering}

The recent observation of the coherent elastic neutrino nuclear scattering, so-called CEvNS, by the COHERENT Collaboration~\cite{COHERENT:2017ipa} has opened a new window to probe extremely low-energy neutrino interactions, a neutrino wave interacting with entire nucleons inside a nucleus.
A neutrino makes a tiny interaction with matter and requires a large target mass for detection. However, physicists have unveiled a new technique to detect neutrinos even with a much smaller detector. A neutrino with an energy above 200 MeV has a wavelength shorter than 1 fm, roughly the size of a nucleon, and thus weakly interacts with an individual nucleon inside a nucleus. On the other hand, a neutrino can interact with all of nucleons in a nucleus when its wave property is pronounced with low energies, below 100 MeV, and thus the probability of neutrino interaction is significantly enhanced. The interest in CEvNS detection comes from not only a sensitive probe of the neutrino wave phenomena in scattering but also a valuable tool of expanding our knowledge of neutrino properties. 


The CEvNS cross section is the largest among neutrino scattering channels for neutrino energies below 100 MeV for most target nuclei. The COHERENT CsI(Na) detector accumulated 306$\pm$20 CEvNS candidate events with a 14.6~kg target mass, consistent with the Standard Model (SM) prediction of 341$\pm$11(theory)$\pm$42(experiment)~\cite{COHERENT:2021xmm}. However, this is insufficient for making efficient searches for new physics because of large statistical and systematic uncertainties. 

To be sensitive to interesting searches and measurements, we propose to develop a CsI(Na) crystal detector with a total of 300 kg target mass, 20 times larger than the COHERENT, and an improved detector design. 
The detector will be able to provide the world-largest CEvNS event rate sufficient for precise measurements and new physics searches that would never be possible without it. 

It expects to provide a complete picture of neutrino weak couplings predicted by the SM by making a precise measurement of the CEvNS cross sections at low momentum transfer. It allows probes of nuclear structure and improved constraints on the value of the weak nuclear charge. The CEvNS process is also useful for probing a yet unknown neutrino properties which can be only explained by the physics beyond the SM (BSM). They include searches for a neutrino magnetic moment, sterile neutrinos, and non-standard neutrino interactions (NSI) mediated by new particles. A CEvNS detector based on the dark-matter detection technique makes it plausible to observe extremely low-energy neutrinos which are undetectable because of their tiny cross-section. Due to the significantly enhanced rate for a lower energy neutrino, the CEvNS effect can be utilized to develop a small-size detector sensitive to neutrino emission in neutron stars and during stellar collapse. The CEvNS detector is also sensitive to solar neutrinos, atmospheric neutrinos, and Diffuse Supernova Neutrino Background which are irreducible backgrounds against direct searches for the dark matter.


   
High target masses of Cs and I atoms provide a substantially large coherence enhancement for a sufficient CEvNS event rate. Both target atom masses are similar and greatly simplify the detector response. The crystal exhibits a high light yield of $\sim$45 photons/keV$_{ee}$ at 662 keV$_{ee}$, with a light emission peaking at 420 nm, 
 The COHERENT experiment measured the light yield of 15.4$\pm$1.2 PE/keV$_{ee}$ using the $^{241}$Am source~\cite{Collar:2019ihs} and achieved an energy threshold of 7 keV$_{\rm nr}$. An absolute energy scale was obtained as 1.2 PE/ keV$_{\rm nr}$~\cite{Barbeau:2021exu} based on the measured quenching factor of $\sim$9\%.     
 
 The detector is named by CICENNS (sounding as "see CENNS") standing for CsI detector for Coherent Elastic Neutrino Nucleus Scattering. It consists of 15 CsI(Na) low-radioactivity crystals, an array of 32 15-cm thick plastic scintillator panels surrounding the target, and sufficient shielding materials against external gamma and neutron backgrounds.  
Each CsI crystal is viewed by two 5-inch super-bialkali (SBA) photomultipler tubes (PMT) at both ends. 
Each of 28 long and 4 short plastic scintillator panels is viewed by two 4-inch PMTs. 
 A total of roughly 13 tonne shielding materials surround the target crystals with copper, contemporary lead and High-Density Polyethylene (HDPE). 
The detector expects to be deployed at CiADS and will provide abundant CEvNS signals to open a new field of exciting measurements.

It is convenient to access the neutrino source of CEvNS at CiADS, located in Huizhou, Guangdong province. A 1 GeV proton beam with a power of 250 kW in the first stage and 2.5 MW ultimately produces continuous neutron and neutrino fluxes on a spallation target which is placed at 13 m underground inside of an experimental hall. 

\begin{figure}[htb!]
\begin{center}
\includegraphics[width=0.5\textwidth]{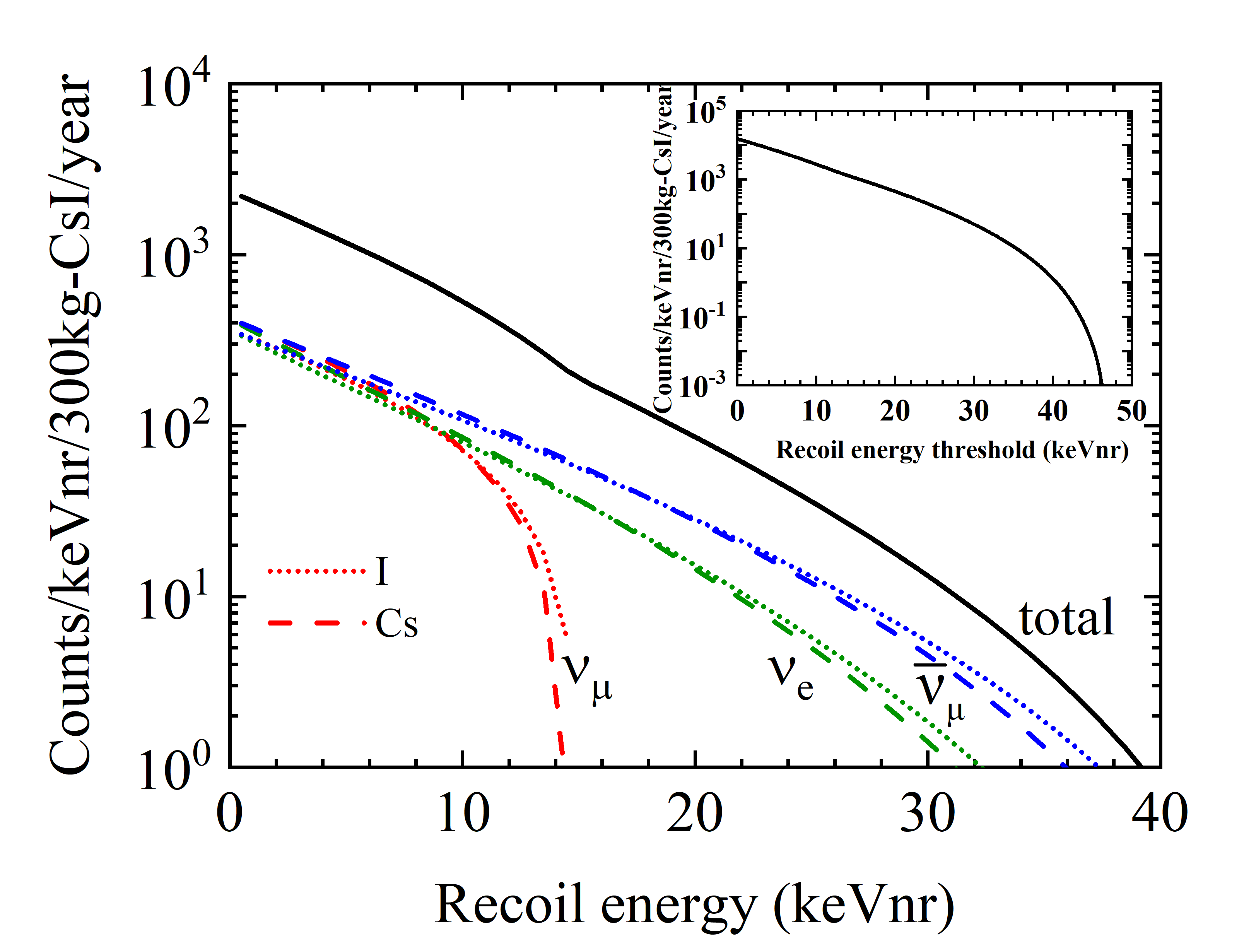}
\end{center}
\caption{Expected nuclear recoil-energy spectrum of CsI(Na) in response to CEvNS at CiADS with a beam power of 250 kW. Note that responses of Cs and I are almost identical. An integrated rate above detector threshold in nuclear recoil energy (keV$_{\rm nr}$) is given in the inset. A realistic threshold is expected to be 5 keV$_{\rm nr}$.}
\label{CsI-NR-E}
\end{figure}

The CICENNS detector expects to be located on the second floor of the experimental hall, approximately 25 m away from the spallation target. There exists a 10 m-concrete shielding structure around the beam target. The neutrino flux at the detector target is expected to be 5.4~$\times 10^6$~/cm$^2$/s, roughly 11.4\% of the flux at COHERENT (4.7~$\times 10^7$~/cm$^2$/s at 20 m from the target for 1.4 MW)~\cite{Barbeau:2021exu}, and will become 5.4~$\times 10^7$~/cm$^2$/s if the beam power is upgraded to 2.5 MW,

The expected nuclear recoil-energy spectrum of CsI(Na) is obtained by taking into account the CEvNS in response to the CiADS neutrino beam and shown in Fig.~\ref{CsI-NR-E}. The nuclear recoil spectra of Cs and I are almost identical due to their similar masses. The maximum nuclear recoil energy is roughly 35 keV$_{\rm nr}$. The expected CEvNS production rate for a 250 kW (2.5 MW) beam is about 2500 (25000) events per 300 kg for a year at the energy threshold of 7 keV$_{\rm nr}$ and 3500 (35000) events at 5 keV$_{\rm nr}$. 

After applying event selection criteria to a generated MC sample, an observed CEvNS signal rate is obtained as a function of nuclear recoil-energy threshold.
The CEvNS observed rate for a 250 kW (2.5 MW) beam is predicted as about 1520 (15200) events per 300 kg for a year at the energy threshold of 7 keV$_{\rm nr}$.

A produced CEvNS event rate at CICENNS is shown in Fig.~\ref{CsI-NR-E} as a function of nuclear recoil-energy threshold. The expected CEvNS production rate per 300 kg for a year is about 2500 (25000) events at the energy threshold of 7 keV$_{\rm nr}$ and 3500 (35000) events at 5 keV$_{\rm nr}$, for a 250 kW (2.5 MW) beam. This would be the world-largest CEvNS data sample. The detection efficiency is 60.5\% above 5 keV$_{\rm nr}$ while it rapidly increases from zero at 1.5 keV$_{\rm nr}$ and becomes 35\% at 3 keV$_{\rm nr}$. An expected CEvNS event rate at CiADS is 1520 (15200) events per 300 kg for a year data-taking at the energy threshold of 7 keV$_{\rm nr}$, for a 250 kW (2.5 MW) beam. The beam correlated background rate is estimated to be about 5\% of a predicted CEvNS rate with an uncertainty of 25\%. 
The steady-state background is estimated to be $4.1 \times 10^5$ per year data-taking at energy threshold of 7 keV$_{\rm nr}$ with its normalization uncertainty of 0.1\%. 

We present expected physics sensitivities of the CICENNS for a 2.5 MW beam here,  
 \newline

\noindent
(a) {Precise measurement of “flavored” CEvNS cross section at low momentum transfer}

\begin{figure}[htb!]
\begin{center}
\includegraphics[width=0.4\textwidth]{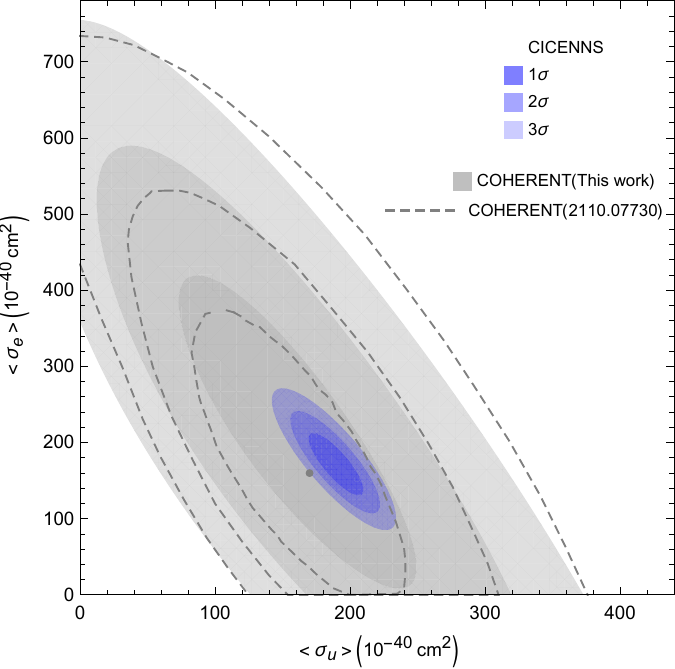}
\end{center}
\caption{Contours for the flavored CEvNS cross section obtained by the CICENNS expectation (blue) in comparison with the COHERENT data (gray)~\cite{COHERENT:2021xmm}. An allowed area expected by the CICENNS is significantly reduced from the COHERENT.}
\label{F-xsection}
\end{figure}

The current uncertainty of the measured CEvNS cross section is +18 or -15\% while that of the SM prediction is 4.8\%. The 300 kg CsI(Na) CICENNS detector will reduce the uncertainty to roughly 5\%, similar to the uncertainty of the nuclear form factor.
Figure~\ref{F-xsection} shows allowed contours in this parameter space, expected from the CICENNS and obtained from the COHERENT.

\vspace{5mm}

\noindent
(b) {Precise measurement of weak couplings at low momentum transfer}

\begin{figure}[htb!]
\begin{center}
\includegraphics[width=0.45\textwidth]{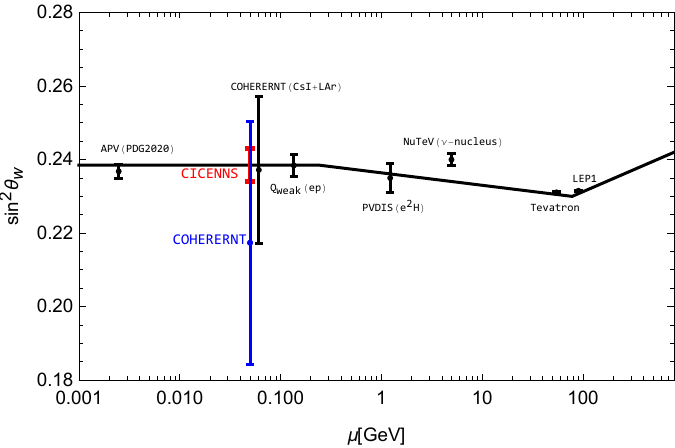}
\end{center}
\caption{Comparison of measured weak mixing angles. The expected error from the CICENNS detector is roughly a percent level while it is $\sim$10\% for the COHERENT.}
\label{weak-coupling} 
\end{figure}

The COHERENT obtained $\sin^2 \theta_{\rm W} = 0.220^{+0.028}_{-0.026}$~\cite{COHERENT:2021xmm} compared to the SM prediction 0.23857(5).
The current uncertainty from the CEvNS measurements is at the $\sim 10\%$ level at Q values of a few tens of MeV/c. The 300 kg CsI(Na) CICENNS detector will be able to make a percent-level ($\sim 1\%$) measurement.  Figure~\ref{weak-coupling} shows measured weak mixing angles and  the CICENNS expected sensitivity. 

\vspace{5mm}

\noindent
(c) {Precise measurement of neutron radius}

The current uncertainty of the measured neutron radius from the CEvNS measurements is at the $\sim$8\% level (1$\sigma$) and the CICENNS detector is expected to provide a measured value of $\sim$1\% level (1$\sigma$) 

\begin{figure}[htb!]
\begin{center}
\includegraphics[width=0.45\textwidth]{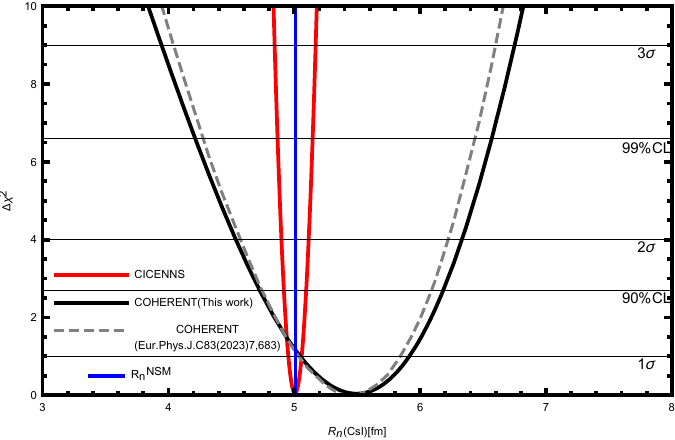}
\end{center}
\caption{Extracted mean radius of neutron distribution (neutron radius) from the observed CEvNS nuclear recoil energy spectrum. The expected error from the CICENNS detector is roughly a percent level while it is $\sim$ 8\% for the COHERENT.}
\label{neutron-radius}
\end{figure}

\vspace{5mm}
\noindent
(d) {Search for non-standard neutrino interactions}

\begin{figure}[htb!]
\begin{center}
\includegraphics[width=0.5\textwidth]{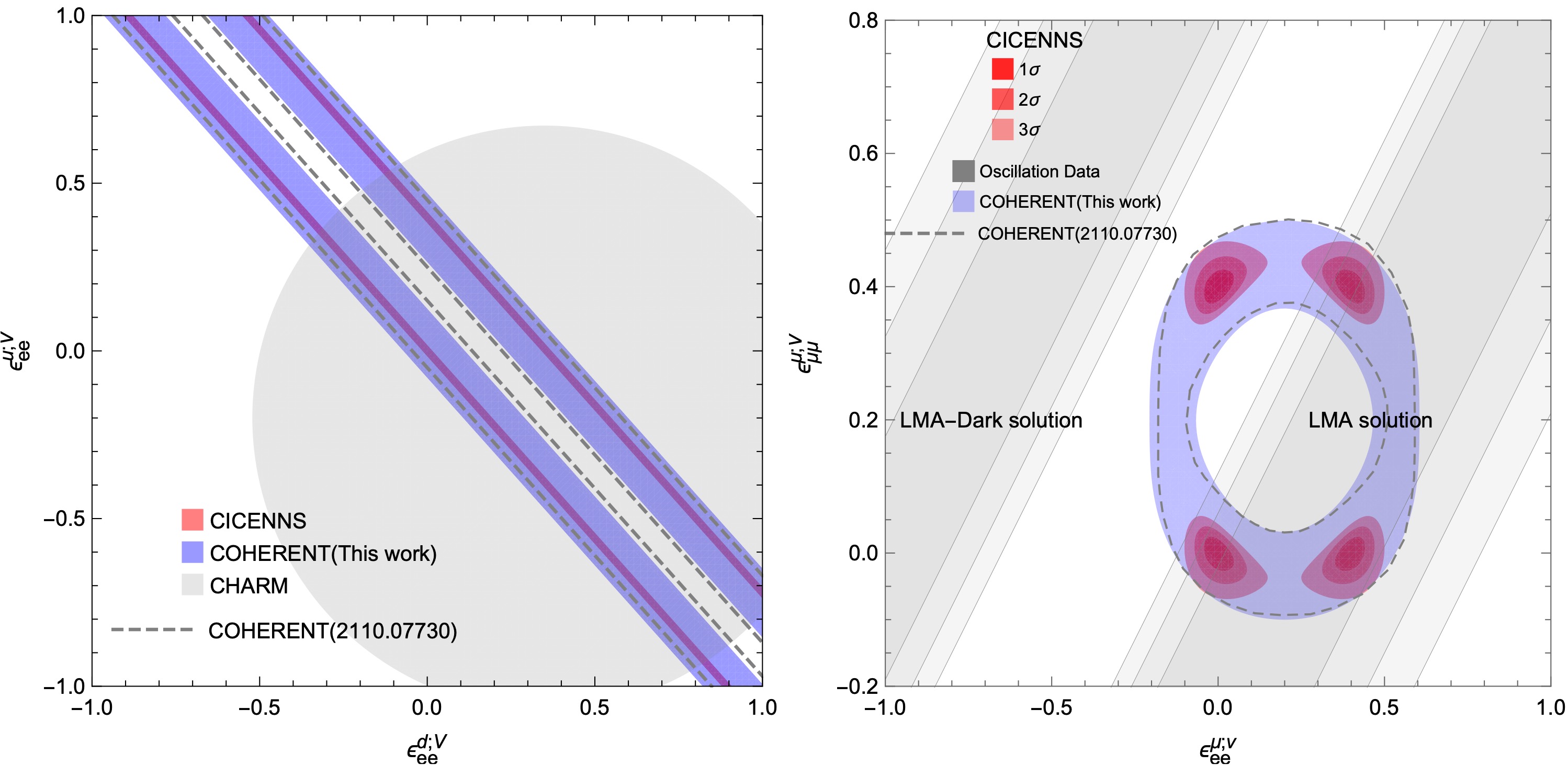}
\end{center}
\caption{Constraints obtained from the COHERENT measurement (blue)\cite{COHERENT:2021xmm} and expected from the CICENNS (red). Left: the 90\% allowed parameter space with $\epsilon^{u}_{ee}$ and $\epsilon^{d}_{ee}$ allowed to float while fixing others at zero. Right: the 1/2/3$\sigma$ contours allowing $\epsilon^{u}_{ee}$ and $\epsilon^{u}_{\mu\mu}$ to float, fixing others to zero. The expected CICENNS constraint is significantly improved relative to the COHERENT. }
\label{epsilon}
\end{figure}

Based on the measured “flavored” CEvNS cross sections, it can provide a sensitive probe of BSM physics such as neutrino-quark vector NSIs, which can affect each neutrino flavor differently. The allowed contour in the $\epsilon^{u}_{ee}$ and $\epsilon^{u}_{\mu\mu}$ parameter space is
shown in Fig.~\ref{epsilon}. 
Due to the sufficiently large CEvNS event rate, the CICENNS search results in significant improvement in the allowed parameter space.

\vspace{5mm}

\noindent
(e) {Search for neutrino magnetic moment}

\begin{figure}[htb!]
\begin{center}
\includegraphics[width=0.5\textwidth]{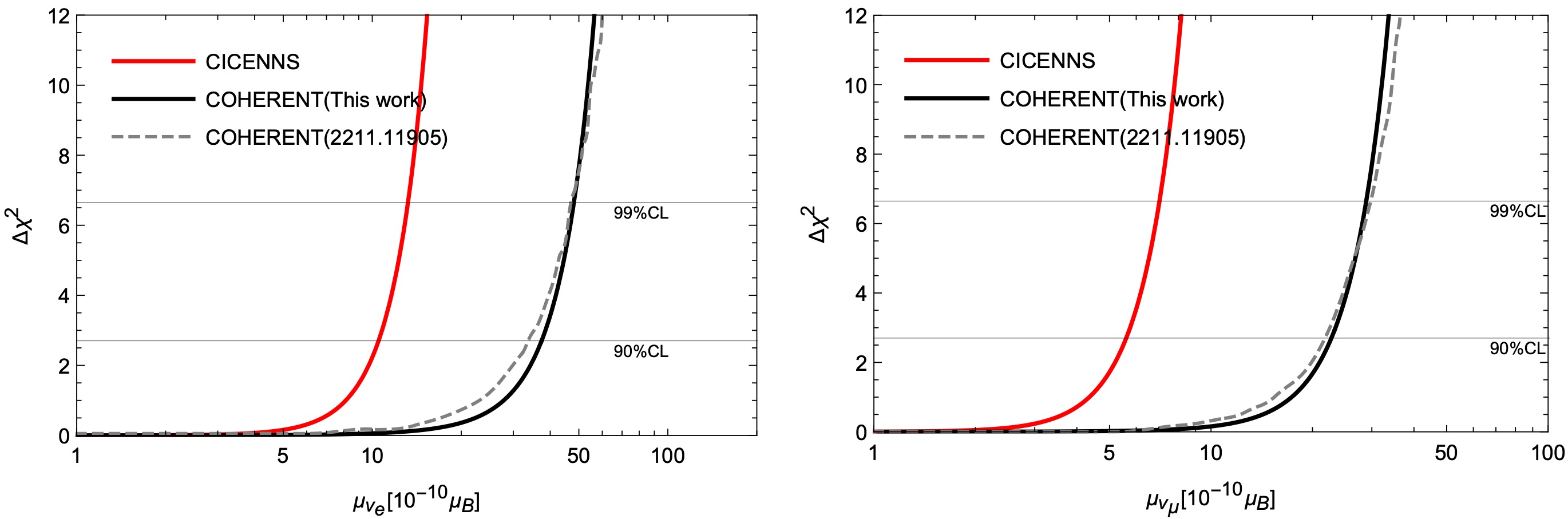}
\end{center}
\caption{Comparison of CICENNS and COHERENT sensitivities on search for neutrino magnetic moment. Left: sensitivity on $\nu_e$ magnetic moment, Right: sensitivity on $\nu_\mu$ magnetic moment.}
\label{NMM}
\end{figure}

A neutrino magnetic moment less than  $10^{-11} \mu_B$ can be measured by the modified the nuclear recoil energy distribution in response to CEvNS. However, an extremely low energy threshold below 1 keV$_{\rm nr}$ and good energy resolution are required for the measurement. The measurement sensitivity of CICENNS in comparison with COHERENT is shown in Fig.~\ref{NMM}.

\subsection{Incoherent Scattering for Hadron physics}

Neutrinos and quarks can undergo weak interaction processes.
When the neutrino energy is sufficiently high, similar to deep inelastic scattering, neutrinos can engage in incoherent processes with nucleons, meaning neutrinos interact with individual nucleons or even the quarks within nucleons. 
Therefore, the nuclear reaction processes of high-energy neutrino beams provide a unique perspective for studying hadron physics. 
The charged $W$ boson weak interaction current can alter the flavor of quarks, endowing neutrino-nucleus scattering with two distinctive features compared to typical electromagnetic and strong interaction processes. 
Firstly, neutrino scattering can reduce the number of final-state hadrons, facilitating the search for new excited states of hyperons~\cite{Wu:2013kla} and $\Xi$ particles. 
Secondly, it enables the detection of the $q\bar{q}$ ($q=u,d,s$) component inside nucleons through neutrino-nucleon scattering.

In conventional hadron reactions, the beams consist of electrons, photons, pions, and kaons. 
Among these, the first three beams do not contain $s$ quark.
Therefore, producing hyperons or $\Xi$ particles by bombarding proton or nuclear targets inevitably involves kaon production. 
Considering the strong decays of hyperons and $\Xi$ excited states, at least three final-state hadrons are produced.
Kaon beams, however, face significant energy limitations, with issues in brightness and stability at high energies. 
In contrast, neutrino beams can generate $s$ quark by altering the $u$ or $d$ quark in nucleons, eliminating the need for associated kaon production. 
Similar to $Kp$ scattering, neutrino beams can directly produce hyperon excited states, thereby reducing the number of final-state hadrons. 
Additionally, the energy range of neutrino beams far exceeds that of kaon beams, making them a novel approach for studying hyperon and $\Xi$ excited states.

Furthermore, due to vacuum fluctuations, nucleons inherently contain $q\bar{q}$ components, but these are challenging to detect through conventional scattering processes. 
(Anti-)Neutrinos, by changing quark flavors, can convert $s\bar{s}$ components into ($s\bar{c}$)$c\bar{s}$. 
Thus, by identifying final states with net quark content of ($uuds\bar{c}$)$uudc\bar{s}$ in (anti)neutrino-proton reactions, it is possible to confirm their origin from (anti)neutrino scattering with ($\bar{s}$)$s$, thereby determining the $s\bar{s}$ component and its distribution in the proton. 
Similarly, the $u\bar{u}$ component can be identified through anti-neutrino scattering with final states of net quark content $uuud\bar{s}$, while the $d\bar{d}$ component can be determined via neutrino scattering with final states of net quark content $uudd\bar{c}$.

\subsection{Three-dimensional PDFs}

The energy of HIAF neutrinos can reach 10 GeV, which is suitable for deep inelastic scattering (DIS) measurements. The DIS of neutrino plays an important role in studying partonic structures of nucleons or nuclei. It provides not only information on the flavor separation which cannot be realized in the charged lepton DIS experiments alone but also on electroweak physics by considering both the neutral-current and charged-current interactions. Measurable quantities in DIS are expressed in terms of parton distribution functions (PDFs) which reveal the longitudinal momentum distributions of partons or the one-dimensional structure of the nucleon. To explore the three-dimensional or the transverse momentum-dependent (TMD) PDFs, one is supposed to consider the semi-inclusive DIS (SIDIS) where a jet or a hadron is also measured in addition to the scattered lepton. Jet can be a direct probe of analyzing properties of the quark intrinsic transverse momentum by utilizing intrinsic asymmetries \cite{Yang:2022xwy,Yang:2023zod}. 
Nevertheless, jet production SIDIS process can not cover the low energy kinematic regions or access the chiral-odd PDFs due to the conservation of the helicity \cite{Yang:2022sbz}, and has difficulties to identify light flavor PDFs even in charged-current scattering, since no hadron is measured or tagged. 
It is therefore to consider the hadron production SIDIS process where the cross section is given by the convolution of TMD PDFs and fragmentation functions~(FFs). In this process, TMD PDFs and FFs are measured simultaneously with flavor identifications. Specially, measuring the dimuon signals in the charged-current SIDIS can provide  promising direct experimental constraints on the strange and anti-strange quark distributions of the nucleon \cite{Olness:2003wz,Gao:2008cu}.

Neutrino DIS also play an important role in studying nuclear effects or nuclear TMD PDFs. In the inclusive process, the formalism of the neutrino nucleus scattering is the same to that of the neutrino nucleon scattering \cite{SajjadAthar:2022pjt} and the target mass effect has been studied in depth \cite{Ruiz:2023ozv}. However, systematic definitions of the nuclear TMD PDFs are still missing. Nuclear TMD PDFs are obtained by decomposing the nuclear two-point correlation function which is defined via nuclear state $|A\rangle$. Both of them need more researches.

\subsection{Three-dimensional GPDs and TDAs of hadrons}

The first moments of Generalized Parton Distributions (GPDs) lead to the hadronic gravitational form factors (FFs), and one can study the proton spin crisis and mass structure, pressure, and shear force distributions of hadrons with the help of gravitational FFs~\cite{Ji:1996ek,Leader:2013jra,Ji:2020ena,Polyakov:2018zvc,Burkert:2023wzr}. Therefore,  the study of GPDs is regarded as one of  the main targets  for the upcoming Electron-Ion Collider (EIC).
There are three golden reaction channels where the hadronic GPDs are accessed: Deeply Virtual Compton Scattering (DVCS), Deeply Virtual Meson Production (DVMP), and Timelike Compton Scattering (TCS). However, one can not distinguish the GPDs of different quark flavors using these  golden channels. To address this issue,  physicists have proposed to use neutrino beams instead of electron beams to study the quark flavor-dependent GPDs, and they are neutrino DVCS and DVMP,
\begin{equation}\label{sl-dis}
 \begin{aligned}
\mathrm{Neutrino \,\, DVCS }: &W(q)+N(p)\to \gamma(q_1)+N_1(p_1),\\
 \mathrm{Neutrino \,\, DVMP }: &W(q)+N(p)\to M(q_1)+N_1(p_1),
 \end{aligned}
\end{equation}
where $M$ is a meson state, $W$ is the  $W$ boson emitted from neutrino, and we use $N$ and $N_1$ to represent  the nucleons in initial and final states, respectively.
Since the cross section of neutrino DVCS is suppressed by an additional electromagnetic coupling constant $\alpha$  compared to the neutrino DVMP, thus, it will be more feasible to measure the latter by experiment~\cite{Pire:2015iza, Pire:2017lfj}.
In these reactions, the energy of the neutrino beam should be larger than 5 GeV so that the QCD factorization condition ($-q^2 \gg -t =-(p-p_1)^2$) is satisfied, and the hadronic GPDs are extracted through the analysis of the measurements at the forward region (small-$t$ region).

 Meanwhile, if we  investigate the backward region (small-$u=(p-q)^2$ region) of these processes, and the nucleon-photon and nucleon-meson transition distribution amplitudes (TDA) can be also extracted from neutrino DVCS and DVMP processes, respectively~\cite{Pire:2021hbl}.  Recently, the scientists at Jefferson Lab (JLab) just  reported the first  measurement of hadronic TDA using DVCS reaction with electron beams~\cite{JeffersonLabFp:2019gpp}. Similarly, one can investigate the TDAs of different quark flavors  by measuring the neutrino DVCS and DVMP processes, which can be complementary to the ongoing measurements at JLab.

\section{Neutron Physics}
\subsection{Bound beta-decay of neutron measurement by microcalorimeter}
The bound beta-decay (BoB) of neutron is a rare decay mode into a hydrogen atom and an anti-neutrino. The state of neutrino can be exactly inferred by measuring the state of hydrogen atom, providing a possible pathway to explore new physics. However, this rare decay mode has not yet been observed so far, since it was predicted in 1947. The challenge in observing this decay is not only that its cross section is extremely low, equivalent to about branching ratio of the order of $10^{-6}$ of the three-body decay, but also that the final-state hydrogen atom is neutral and has extremely low kinetic energy, which cannot be effectively detected. Here we propose a microcalorimeter-based scheme for measuring the kinetic energies of hydrogen atoms produced from BoB of ultracold neutrons, which has a great advantage in terms of accuracy of the energy measurement.  

For the BoB mode, the conservation of momentum and energy during the decay leads to the energy carried by the antineutrino is constant, while the kinetic energy of the hydrogen atom is also constant (325.7 eV),making it easy to be identify. Since the hydrogen atom in the $n$s ($\geq 2$) state after the two-body decay can be captured by the detector, it will de-excite to the 1s state very quickly, transferring its thermal energy 10.2 eV more to the detector. Thus, the energy(kinetic energy plus potential energy) spectra of the hydrogen atoms on the detector would be a multiple line structure dominated by 325.7 eV and 335.9 eV. A microcalorimeter is a detector sensitive to heat release and has an extremely high energy resolution of the order of eV with no entrance window (dead layer) and offers the possibility of a wide range of available absorption materials~\cite{NSTTES2022}. The high energy resolution of the order of eV allows to resolve the signal from emerging hydrogen atoms, which produce a narrow energy peak at 325.7 eV. Hydrogen atoms, electrons and protons can all produce heat signals in the detector. To get higher peak to background ratio, 
a 1kV anode will be used to remove Protons. The 2s H atoms will de-excited to 1s in strong electric field, only 1s peak is enough for aim to observe BOB, We need find a way to remove Protons without de-excite 2s H.

\subsection{Parity-Violating Nuclear Forces from Neutron Capture Measurement}

The emergence of parity-violating hadron-hadron interactions within the Standard Model remains unsolved and unverified experimentally. Experimentally, the only way to study strangeness-conserving weak interactions between hadrons is to measure parity violation in nucleon-nucleon interactions \cite{Ramsey-Musolf:2006vfz, Haxton:2013aca}. Due to the weakness of the interaction compared to the strong and electromagnetic forces, parity-violating signals in nucleon-nucleon processes are extremely small and obscured by strong interactions. To date, N-N parity violation has only been observed in a few reaction channels. High-precision measurements using high-intensity polarized neutrons interacting with hydrogen and light nuclei could reveal clear parity-violating signals, advancing our understanding of hadronic properties and completing the Standard Model’s description of weak interactions between hadrons.

Parity violation in nucleon-nucleon interactions is typically determined through analyzing powers in single-polarization experiments, angular asymmetries of final-state particles, or polarization of emitted photons. These observables are minuscule, on the order of $o(10^{-7})$. Parity violation of significance has so far only been observed in proton-proton and proton-helium elastic scattering \cite{Eversheim:1991tg, Nagle:1978yc, Kistryn:1987tq, TRIUMFE497:2002sue, Lang:1985jv}. Recently, a weak parity-violating signal was detected in the ${\rm n-^3He}$ system at Oak Ridge National Laboratory’s neutron source \cite{n3He:2020zwd}. Overall, high-precision experimental data on nucleon parity violation remain scarce, particularly for neutron capture processes (${\rm n-p, n-d, n-^3He}$) \cite{n3He:2020zwd, NPDGamma:2018vhh, Cavaignac:1977uk, Gericke:2011zz, Alberi:1988fd}, where no clear parity-violating effects have been observed.

Over the next 5–10 years, CiADS will provide the highest-power proton beams in the hundreds of MeV energy range. Its high-power target terminal can generate intense neutron sources, enabling high-precision parity-violation asymmetry measurements in cold neutron capture processes. We propose using CiADS’s high-flux neutrons to search for parity violation in polarized cold neutron capture by hydrogen, deuterium, and ${\rm ^3He}$. In the first phase, we aim to measure the angular asymmetry of protons in the ${\rm ^3He(\vec{n},p)^3H}$ reaction with $o(10^{-8})$ precision level. 

Detecting nonzero parity-violating signals is highly challenging. First, sufficient neutron flux ($>10^{16}$ neutrons) is required to achieve statistical significance. Second, systematic errors must be rigorously controlled to eliminate spurious signals. The key technical challenges for the future experiment include: (a) high-intensity, temporally structured cold neutron beams; (b) real-time neutron polarization techniques; (c) stable, high-precision ${\rm ^3He}$-gas target and multi-wire proportional chamber integrated system for neutron capture process and detection of final proton.

\section{Symmetries and Quantum Effects in Nuclear Physics}

\subsection{Test of fundamental symmetries at nuclear energy scales}


At the high-intensity frontier, experimental searches for key processes and observables—such as neutrinoless double-beta decay and electric dipole moments (EDMs) of fundamental or composite particles—aim to test fundamental symmetries with unprecedented precision. In addition, nuclear superallowed Fermi transitions serve as a high-precision probe for determining the CKM matrix element $V_{\rm ud}$ and testing its unitarity. In all these studies, accurate knowledge of  nuclear matrix elements is essential to connect experimental signatures with the potential new physics. High-precision measurements of the properties of related atomic nuclei necessitate advanced facilities like HIAF, which can produce intense, high-quality beams required for such studies. The data obtained from HIAF will not only enhance our understanding of fundamental symmetries in nature but also offer stringent tests for nuclear models.

\subsubsection{Neutrinoless double-beta decay}


Neutrinoless double-beta ($0\nu\beta\beta$) decay is not only a crucial experiment for testing whether neutrinos are Majorana particles and whether lepton number is conserved, but it is also expected to provide information on the absolute neutrino mass. As a result, it has long been at the forefront of research in particle and nuclear physics. Several national and regional scientific strategic plans in China~\cite{China:NLDBD}, the United States~\cite{US:LRP}, and Europe~\cite{EU:LRP} have listed it as a priority research direction. In China, active research is being conducted on the $0\nu\beta\beta$ decay of five key candidate isotopes, including  $^{76}$Ge on CDEX~\cite{CDEX:2024}, $^{82}$Se on N$\nu$DEx~\cite{NnDEx-100:2024}, $^{100}$Mo on CUPID-China~\cite{Ma:2024}, $^{130}$Te on JUNO~\cite{Cao:2020}, and $^{136}$Xe on PandaX~\cite{PandaX:2024}.


The nuclear matrix element (NME) $M^{0\nu}$ serves as the critical link between the  half-life $T^{0\nu}_{1/2}$ of $0\nu\beta\beta$ decay and  the effective electron neutrino mass $\langle m_{\beta\beta}\rangle$ in the following way
\begin{equation}
\label{half-life}
 |\langle m_{\beta\beta}\rangle|
=\left[\dfrac{m^2_e}{g^4_A G_{0\nu}T^{0\nu}_{1/2}
 \left\lvert M^{0\nu}\right\rvert^2}\right]^{1/2},
\end{equation}
where $m_e\simeq0.511$ MeV is electron mass. The phase-space factor $G_{0\nu}\approx 10^{-14}$ yr$^{-1}$ can be evaluated rather precisely.  The precise calculation of the NMEs helps guide experimental design, including optimizing the choice of candidate nuclei and experimental setups, thereby enhancing the detection prospects. However, the NME cannot be directly measured in experiments and must be determined through theoretical calculations. 

For a long time, different phenomenological nuclear models have produced significantly varying NME values, with discrepancies reaching up to a factor of three or more~\cite{Yao:2022PPNP,Agostini:2023}, leading to considerable uncertainty in the theoretical interpretation of experimental results. These discrepancies mainly arise from fundamental differences in the treatment of nuclear forces and the truncation of model spaces used to describe nuclear wave functions, making the reduction of model dependence in NME calculations a major challenge.  Recent advancements in ab initio nuclear-structure methods have shown promise in providing consistent NME values with theoretical uncertainties for lighter double-beta decay candidates~\cite{Yao:2020PRL,Belley:2021,Novario:2021}. However, the uncertainties—dominated by many-body truncations—remain substantial~\cite{Belley:2024PRL}. Promising ways to mitigate this problem include refining the approximations of nuclear models and exploring correlations between the NMEs and experimental observables from low-energy nuclear structure experiments. These observables include the double Gamow-Teller strength~\cite{Shimizu:2018,Wang:2024}—a double spin-isospin flip excitation mode that can be probed through high-energy heavy-ion double-charge-exchange  reactions~\cite{NUMEN:2023PPNP}—as well as nucleon transfer reactions~\cite{Brown:2014}, double-gamma transitions~\cite{Romeo:2025}, high-energy heavy-ion collisions~\cite{Li:2025}, and properties of nuclear low-lying states~\cite{Zhang:2024}. {Therefore, precision measurements of nuclear structure properties and reaction rates—such as proton-removal cross sections from ($d$, $^3$He) reactions~\cite{Kay:2009} and two-neutron transfer reactions like ($p$, $t$)~\cite{Rebeiro:2020}—at HIAF will contribute to refining the nuclear models used to calculate NMEs for neutrinoless double-beta decay.}

\subsubsection{Atomic EDMs, nuclear Schiff moments and PT-odd nuclear forces}

In 1963, Schiff demonstrated that for an atom with a point-like nucleus and nonrelativistic electrons, the effect of any nonzero nuclear electric dipole moment (EDM) would be completely screened by the surrounding electrons, irrespective of the external potential, resulting in a net atomic EDM of zero~\cite{Schiff:1963}. However, this screening is incomplete when the finite size of the nucleus or the magnetic interactions arising from the relativistic effect is considered. In such cases, atoms can acquire nonzero EDMs due to a residual effect, which is induced by what later became known as the Schiff moment.

The PT-violating electrostatic interaction between atomic electrons and the nuclear Schiff moment mixes atomic states of opposite parity, inducing a T-violating EDM in the atomic ground state. Thus, the atomic EDM $d_A$ depends on the Schiff moment of the nucleus, which arises from PT-violating nuclear potential $\hat{V}_{PT}$. In the first-order perturbation approximation, the nuclear Schiff moment $S$ is given by  
\begin{equation}
\label{eq:Schiff_moment}
S\simeq
\sum_{i\neq 0}\cfrac{\langle \Phi_0|\hat{S}_z|\Phi_i\rangle\langle \Phi_i|\hat{V}_{PT}|\Phi_0\rangle}{E_0-E_i}+c.c. 
\end{equation}
The PT-violating nuclear potential $\hat{V}_{PT}$ is induced from the CP-violation sources at high energy scales and can be parameterized at nuclear energy scales by introducing the  PT-violating $N\pi$ coupling vertex at the leading-order of chiral expansion~\cite{deVries:2020}
 \begin{eqnarray}
   \mathcal{L}_{PVTV} &=& \bar g^{(0)}_{\pi NN}\bar N N\vec{\tau}\cdot\vec{\pi}
   + \bar g^{(1)}_{\pi NN}\bar N N \pi_z\nonumber\\
   &&
   + \bar g^{(2)}_{\pi NN}\bar N N(3\tau_z\pi_z-\vec{\tau}\cdot\vec{\pi}) 
\end{eqnarray} 
which consists of iso-scalar, iso-vector and iso-tensor terms. The $\hat{V}_{PT}$ is a scalar operator that changes parity and $\hat{S}_z$ is a rank-one tensor operator. The intermediate excited states $|\Phi_i\rangle$ must have the same total angular momentum as the ground state but opposite parity, and the angular momentum must be nonzero. Based on these considerations, atoms with odd-mass nuclei are typically chosen in searches for a permanent EDM.  


By measuring the atomic EDM, constraints can be placed on the magnitude of the Schiff moment, which in turn limits the low-energy coupling constants (LECs)  $\bar{g}^{(i)}_{\pi NN}$. The nuclear Schiff moment can ultimately be expressed as  
\begin{equation}
S=g_{\pi NN}(a_0 \bar{g}^{(0)}_{\pi NN}+a_1 \bar g^{(1)}_{\pi NN}+a_2 \bar g^{(2)}_{\pi NN}),
\end{equation} 
where the coefficients $a_{i=0,1,2}$ encode  all nuclear structure information. It is evident that an accurate determination of $a_i$ is crucial for precisely constraining $\bar{g}^{(i)}_{\pi NN}$. However, accurately determining $a_i$ remains a major theoretical challenge, as these coefficients are highly sensitive to the excitation energies and wavefunctions of parity doublet states.  

Previous studies have shown that the coefficients $a_{0,1,2}$ of the nuclear Schiff moment for experimentally relevant nuclei $^{129}$Xe, $^{171}$Yb, $^{199}$Hg, and $^{225}$Ra) exhibit significant variations across different nuclear models, spanning two to three orders of magnitude and sometimes even differing in sign~\cite{Engel:2013,Zhou:2023,Engel:2025}. This large uncertainty in $a_i$  introduces a considerable ambiguity in the nuclear Schiff moment, thereby affecting the allowed values of the LECs characterizing PT-violating nucleon-nucleon interactions.

For weakly deformed nuclei such as $^{199}$Hg, high-lying excited states may contribute significantly to the Schiff moment, as the quenching effect from the energy denominator is relatively moderate. This phenomenon has been observed in both random-phase approximation (RPA)\cite{Ban:2010} and shell-model\cite{Yanase:2020} studies. However, the existence of such states in $^{199}$Hg remains to be confirmed experimentally. Moreover, recent studies have revealed a strong correlation between the intrinsic Schiff moment of odd-mass nuclei and the octupole moment of neighboring even-even nuclei~\cite{Dobaczewski:2018PRL}.   This correlation holds promise for significantly reducing systematic uncertainties among different theoretical models. Therefore, experimental measurements of structural properties of $^{129}$Xe, $^{171}$Yb, $^{199}$Hg, and $^{225}$Ra  are crucial. They can help validate nuclear structure models and ultimately lead to more reliable estimates of the nuclear Schiff moment.

$^{225}$Ra is particularly interesting due to its enhanced Schiff moment, which arises from its large octupole deformation and the small energy splitting between positive- and negative-parity states. This enhancement is predicted to make $^{225}$Ra up to a thousand times more sensitive than $^{199}$Hg in searches for electric dipole moments (EDMs). Efforts to search for EDMs using $^{225}$Ra began over a decade ago, but progress has been limited due to the short half-life and radioactivity of the $^{225}$Ra nucleus. In recent years, an EDM measurement setup has been developed by Lu’s group at USTC and has already been successfully used in the EDM search for $^{171}$Yb. $^{225}$Ra is expected to be the next target, and related preparations for EDM measurements are currently underway.

Most importantly, the recent success in laser spectroscopy of the radioactive molecule RaF has opened a new window for EDM searches in heavy, deformed nuclear systems \cite{GarciaRuiz:2019kwh}. This approach takes full advantage of both the $10^3$-fold enhancement of the Schiff moment in octupole-deformed $^{225}$Ra and the $10^4$-fold enhancement of the effective internal electric field provided by the molecular structure. Together, these offer unprecedented opportunities to search for CP violation in $^{225}$Ra within a designed molecular system \cite{Arrowsmith-Kron:2023hcr}. Such explorations of radioactive molecules are also being planned in China \cite{Hu:2025izj}, which may greatly benefit from the Huizhou Large-Scale Scientific Facilities once the ISOL system is implemented in the near future.

\subsubsection{Superallowed Fermi transitions}
The $0^+ \to 0^+$ superallowed  Fermi transitions are a special class of nuclear beta decays where a nucleus undergoes a pure vector transition $\Delta J=0, \Delta T=0$ between isospin analog states (IAS), differing only in isospin projection. Precision measurements of superallowed Fermi transitions are a critical tool to search for physics beyond the
standard model, including both in the extraction of  
the CKM matrix element $V_{\rm ud}$   and search of new physics~\cite{Hayen:2024,Gorchtein:2024}.

In combination of the measured superallowed beta transitions of nuclei from $^{10}$C to $^{74}$Rb with theoretical calculations, one finds the experimental $\mathcal{F}t$ value of individual nucleus~\cite{Hardy:1975,Hardy:2009}  
\begin{equation}
\label{eq:corrections}
\mathcal{F}t = ft (1 + \delta'_R)(1 + \delta_{NS} - \delta_C),
\end{equation}
and the nucleus-independent corrected $\overline{\mathcal{F}t}$ value. In the above formula,  $ft$ is the product of the measured half-life $t$ and the statistical rate function $f$, $\delta'_R$ is the transition-dependent part of the radiative correction, $\delta_{NS}$ is the nucleus-specific component of the radiative correction, and $\delta_C$ is the isospin symmetry-breaking correction (on the order of $Z^2\alpha^2$~\cite{Towner:1973})
\begin{equation}
M_F^2= 2(1-\delta_C),
\end{equation}
that modifies the Fermi matrix element $M_F$ as a result of Coulomb and nuclear charge-dependent forces. The $\delta_C$ value typically ranges from 0.1\% to 1\%, with its magnitude depending on the choice of nuclear model~\cite{Gorchtein:2024}. It is noteworthy that if the conserved vector current (CVC) is satisfied, the $\mathcal{F}t$ value for the pure Fermi superallowed transitions should be a nucleus-independent constant. 

The value of $V_{\rm ud}$ is determined from the $\overline{\mathcal{F}t}$ value~\cite{Hardy:2009} 
\begin{equation}
 V_{ud}^2 = \frac{K}{2 G_F^2 \overline{\mathcal{F}t} (1 + \Delta_R^V)},
\end{equation}
 where $K/(\hbar c)^6 = 8120.27648(26) \times 10^{-10}$ GeV$^{-4}$s is a fundamental constant, $G_F/(\hbar c)^3= 1.1663788(6) \times 10^{-5}$ GeV$^{-2}$ is the Fermi coupling constant, and $\Delta_R^V$ is the nucleus-independent radioactive correction. 
 
 Thus, a precise determination of nuclear structure dependent corrections in (\ref{eq:corrections}) with reduced uncertainties in the superallowed Fermi transitions is necessary to ensure an accurate test of CKM unitarity and the Standard Model. Considerable effort is ongoing to improve these calculations by combining new experimental measurements with advances in modeling the transitions with operators from chiral EFT and nuclear wave functions from nuclear ab initio theories~\cite{Stroberg:2021,Cirigliano:2024PRL,Cirigliano:2024,Gennari:2025}. The HIAF is poised to measure key superallowed Fermi transitions, providing data that will serve as a critical test for nuclear theories related to these transitions.

\subsection{Quantum Effects in Nuclear Physics}


Historically, the exploration of quantum entanglement in nuclear systems dates back to the pioneering work of Lamehi-Rachti and Mittig in 1976 \cite{Lamehi-Rachti:1976wey}, who conducted the first nuclear Bell test by analyzing spin correlations in low-energy proton-proton scattering experiments. Remarkably ahead of its time, this groundbreaking work garnered little attention from the nuclear physics community for the following three decades. A resurgence occurred in 2006 with another nuclear Bell test experiment by Sakai \emph{et al.}\ \cite{PhysRevLett.97.150405}, which introduced critical improvements over the 1976 study and established a modern framework for probing proton spin correlations. Over the next fifteen years, theoretical studies on the entanglement properties of atomic nuclei appear occasionally in the literature. These early efforts gradually captured the attention of the broader nuclear physics community and, along with contemporary efforts of studying nuclear physics on quantum computers, have coalesced into the emerging quantum frontier of low-energy nuclear physics. 

Recent theoretical advances have explored the entanglement properties inherent to nuclear forces, structures, and dynamics. For instance, studies in Refs.\ \cite{Beane:2018oxh,Bai:2022hfv,Bai:2023tey,Miller:2023ujx,Kirchner:2023dvg,Nan:2024byp} quantify the spin entanglement of nuclear forces between nucleons and light nuclei, revealing novel connections between extreme values of quantum correlations and the emergence of low-energy nuclear symmetries, such as Wigner SU(4) and nonrelativistic conformal symmetries. These findings suggest that entanglement may act as a hidden organizing principle underlying nuclear phenomena. In parallel, Refs.\ \cite{Kruppa:2020rfa,Johnson:2022mzk,Perez-Obiol:2023wdz,Gorton:2024hbb} investigate the entanglement structure of nuclear shell models, focusing on ground-state configurations. These works identify remarkably low entanglement between proton and neutron wave function components, and inspires a new approximation ansatz for constructing shell-model wave functions.  Moreover, Refs.\ \cite{Pazy:2022mmg,Gu:2023aoc} explore the scaling behavior of orbital entanglement using generalized contact theory and coupled-cluster methods. Intriguingly, these studies identify a volume-law scaling of entanglement entropy in nuclear systems, contrasting the area-law scaling dominant in condensed matter systems and black hole thermodynamics. This distinction hints at unique quantum information dynamics in nuclei, potentially tied to their finite, strongly interacting nature. Furthermore, pioneering work has extended entanglement analysis to diverse nuclear phenomena, including halo nuclei \cite{Robin:2020aeh}, multi-nucleon transfer reactions \cite{Li:2024jdb}, dynamical fission processes \cite{Qiang:2024syr}, cluster states \cite{Kanada-Enyo:2015ncq}, collective motions \cite{Chen:2024vvd}, nucleon-nucleon short-range correlations \cite{Kou:2023knx}, and two-proton decays \cite{Oishi:2024quz}. 
Note that nuclear fission exhibits an unique example of many-body quantum entanglement in strongly interacting systems due to the persistent non-localization of wave functions in the fast splitting process \cite{Qiang:2024syr}. The inclusion of quantum entanglement in nuclear fission models would also be 
a great opportunity for quantitative descriptions of the distributions of multiple correlated fission observables.  These studies position entanglement as a unifying framework for deciphering nuclear systems, from microscopic interactions to macroscopic emergent behaviors. Last but not least, recent attempts to connect these theoretical results to experimental studies could bring new opportunities to facilities like HIAF and other advanced laboratories worldwide \cite{Bai:2023hrz}.

In HIAF, nuclear entanglement can be systematically explored via the following steps: (1) Generate high-intensity, maximally spin-entangled proton beams using proton-proton scattering. (2) Establish state-of-the-art Bell tests at the mega-electron-volt scale using these entangled proton beams. (3) Reconstruct one- and two-proton spin wave functions experimentally via quantum state tomography. (4) Measure quantum decoherence of spin-entangled proton pairs propagating through material media. (5) Probe polarization observables using spin-entangled proton beams. (6) Characterize entanglement dynamics in direct and compound nuclear reactions with these beams. (7) Pioneer advanced proton imaging techniques using entangled beams. A comprehensive analysis of these steps will be presented in a forthcoming publication.
    
\section{Vortex Beam and Its Physics}

Vortex(twisted) beams are a new type of particle beams characterized by quantized longitudinal orbital angular momentum (OAM). More than thirty years after the first generation of optical vortices, vortex beams have evolved into a broad and dynamic field, spanning laser optics, quantum information, chemistry, astrophysics, atomic physics, nuclear and particle physics ~\cite{shen2019optical,ivanov2022promises,zou2023recent}. 

The first vortex electron beam was generated in 2010 using electron microscopes \cite{uchida2010generation}, and vortex neutrons were obtained from the cold neutron source in 2015 \cite{clark2015controlling}, and more recently in 2021, vortex atoms of non-relativistic helium were successfully generated \cite{luski2021vortex}. The ability to generate various vortex particles is of great significance, offering novel insights into a wide range of applications particularly in accelerator physics, atomic physics, as well as nuclear and particle physics.

\subsection{Accelerator physics}
The study of vortex particles has been severely constrained in the low-energy region by current experiments. Vortex particle beams have not yet been successfully generated in particle accelerators, and the production of high-energy vortex beams remains a significant challenge. 

There are two main possible approaches to achieving high-energy vortex particle beams: (1) First, applying phase modulation (such as spiral phase plates or holographic gratings) to endow low-energy electron beams with specific phases and intrinsic OAM, followed by accelerating the vortex electron beam to a high-energy state using particle accelerators; (2) Accelerating electrons to achieve a high-energy state first, then using phase modulation or specific scattering to obtain OAM. Both approaches present significant theoretical and technical challenges. In particular, in the second approach, ensuring coherence in high-energy states is difficult, and the requirements for phase modulation materials increase significantly with energy. Vortex particles exhibit unique dynamics in electromagnetic fields within accelerators. However, it has been demonstrated that the manipulation of relativistic twisted charged particles using standard linear accelerators is promising \cite{silenko2017manipulating,silenko2018relativistic,silenko2019electric,zou2021general,karlovets2020quantumbusch,karlovets2021quantumbusch,baturin2022evolution}. In comparison, the first approach seems technically viable. However, methods for effectively manipulating or accelerating vortex particles in modern particle accelerators have not been fully understood. Therefore, a comprehensive study of beam dynamics for vortex particles in accelerators has become an urgent priority.

\subsection{Atomic physics}
Vortex beams introduce novel effects in atomic and molecular systems, significantly altering OAM transfer, selection rules, and collision properties. 

In photoionization, vortex photons impart distinct angular momentum characteristics, atoms on the optical axis yield photoelectrons with well-defined OAM, while off-axis displacement induces finite OAM dispersion~\cite{Pavlov_202409}. Studying atom-light interactions, involving twisted light absorption, can reveal novel interference effects~\cite{Ivanov_202108}. Additionally, OAM carrying beams enable to control tunnel ionization in atoms and molecules. The ionization process is manipulated by OAM and phase singularity ~\cite{Begin_202503}. Vortex photons can modify transition selection rules~\cite{ivanov2019fate}. While OAM couples to molecular in electric dipole transitions, it directly influences internal electron states in quadrupole transitions~\cite{Babiker_2002}. The optical quadrupole interaction in atoms is significantly enhanced when interacting with an optical vortex near resonance, leading to notable mechanical effects on translational and rotational motion in atoms ~\cite{Lembessis_201302}. It was shown in ~\cite{Maslov_202409} that structured light carrying OAM can induce and enhance molecular transitions that are forbidden for conventional light. 

The scattering of twisted electrons on atoms differs significantly from that of plane-wave electrons. The standard Born formula for scattering off a potential field has been extended in~\cite{Karlovets_201511} to twisted electron wave packets. OAM-dependent angular distributions reveal sensitivity to beam kinematics and target composition~\cite{Karlovets_201705}. Scattering from diatomic molecules reveals Young-type interference patterns, demonstrating how vortex beam phase structures and intensity inhomogeneities strongly affect scattering distributions~\cite{Maiorova_201810}. Bremsstrahlung emission by vortex electrons in strong nuclear fields exhibits distinct angular and polarization signatures, directly linked to incident electron OAM~\cite{Groshev_202001}.
Vortex atoms have also been experimentally generated recently~\cite{Luski_202109}, theoretical studies further indicate that the distribution of emitted photons for vortex atom can be altered due to the modified nuclear recoil effects~\cite{Maslennikov_202405}.

\subsection{Nuclear physics}
Vortex beams open new avenues for nuclear physics research. Their unique properties modify processes such as isomer depletion, giant resonance excitation, neutron scattering, and neutron decay.

A protocol has been developed in~\cite{Wu_202204} to achieve the external control of the isomeric nuclear decay by using electron vortex beams. Theoretical calculations for $^{93m}$Mo indicate that these beams can enhance isomer depletion by four orders of magnitude compared to spontaneous decay. Traditional photonuclear reactions primarily excite giant dipole resonances, making the measurement of isovector giant resonances with higher multipolarities a great challenge. It was demonstrated in~\cite{Lu_202311} that vortex $\gamma$ photons enable the selective excitation of giant resonances in even-even nuclei. By imposing angular momentum constraints on electromagnetic transitions, these photons allow for the isolation of specific multipolarities. Angular momentum (AM) transfer in giant resonances (GRs) excited by vortex electrons was also recently analyzed in~\cite{Lu_202502}. Vortex electrons can be used to extract GR transition strength as in the plane-wave case. Moreover, relativistic vortex electrons with larger orbital angular momentum can be generated in the process. This offers a new approach for nuclear structure research and the generation of vortex particles.

The process of elastic scattering of neutrons by nuclei at small scattering angles may led to spin asymmetries in the cross section of the scattered neutrons. When twisted neutrons are used the angular distributions of the scattered neutrons are modified, and shows a dependence on longitudinal polarization. The corresponding spin asymmetries are argued to be measurable at existing neutron facilities~\cite{Afanasev_201911}. A theoretical formalism is then developed in~\cite{Afanasev_202105} for scattering of twisted neutrons by nuclei in a kinematic regime where interference between the Coulomb interaction and the strong interaction is essential. Novel observable effects in the scattering cross section, spin asymmetries, and polarization of the scattered neutrons are revealed. Moreover, recent studies on vortex neutron~\cite{Kou_202503, Pavlov_202502} have shown new decay characteristics distinct from those of plane-wave neutrons. The spectral-angular distributions of emitted protons in vortex neutron decay are highly sensitive to the structure of the neutron wave packet. This sensitivity can be used to extract the distinctive features of non-plane-wave neutron states. 

\subsection{Particle physics}
One of the promising applications of vortex particles is their use as an alternative to spin-polarized beams in high-energy collisions. Vortex particles can be utilized to study strong interactions, probe hadron spin, and investigate the parton structure of hadrons~\cite{ivanov2022promises}. High-energy vortex electrons can serve as probes for studying proton structure, where the angular momentum of vortex electrons is transferred to quarks or gluons, enabling new investigations of proton spin. The study of elastic and inelastic scattering of a lepton on a hadron $e_{tw}p_{tw}\rightarrow e’X$, where both particles carry intrinsic orbital angular momentum, may provide a novel approach to analyzing hadron form factors and open new possibilities for quantum tomography of hadron structure. Furthermore, the investigation of deep inelastic scattering in both inclusive and exclusive regimes using vortex particles may lead to the identification of new structure functions that describe the distribution of partons within a proton and their connection to the standard formalism.

Elastic scattering of aligned vortex electron beams enables direct experimental access to the Coulomb phase, which has long been studied theoretically but never measured \cite{Ivanov_201610}. Processes with two twisted particles give access to observables which are difficult or impossible to probe in the usual plane wave collisions. It has been shown in~\cite{Ivanov_202005_a,Ivanov_202005_b} that resonance production in twisted photon collisions or twisted $e^{+}e^{-}$ annihilation serves as a novel probe of spin- and parity-sensitive observables in fully inclusive cross sections with unpolarized initial particles. Notably, it enables the production of nearly 100\% polarized vector mesons in unpolarized twisted $e^{+}e^{-}$ annihilation. This process also exhibits several surprising kinematic features~\cite{Ivanov_202001}, even in the context of spinless particle annihilation. To demonstrate these remarkable features, a key challenge lies in the generation and detection of twisted particles in the GeV energy range. A new diagnostic method utilizing the superkick effect was proposed recently~\cite{Li_202412, Liu_202503} to detect the presence of a phase vortex. A proof-of-principle experiment with vortex electrons was proposed with existing technology, the realization will also constitute the first observation of the superkick effect.

Vortex particle beams exhibit broad application prospects in the investigation of high-precision problems in atomic, nuclear, and particle physics. The ability to generate high-energy vortex beams is a prerequisite for such studies and represents a key development direction for high-precision experiments at the Huizhou facility. Vortex particle accelerators are among the effective approaches to address this challenge. The feasibility of producing vortex charged particles in linear accelerators by optimizing the magnetic-immersed cathode and stripper foil configuration has been theoretically demonstrated by D. Karlovets and his collaborators \cite{karlovets2020quantumbusch, karlovets2021quantumbusch}. Based on this theoretical framework, the team is currently collaborating with researchers at the Joint Institute for Nuclear Research (JINR) in Dubna, aiming to develop a vortex electron source at the LINAC-200 accelerator to generate vortex electron beams with energies up to 200 MeV.

Similarly, we are planning to conduct research on the generation of vortex ion beams at Huizhou based on the magnetic immersion method. Preliminary results further suggest that, for generating high-quality vortex ion beams, magnetic-immersed stripping should be performed when the ion energy is close to or exceeds 1 MeV/u. For negative hydrogen beams, this corresponds to the output energy of the radio-frequency quadrupole (RFQ) section in the front end of an ion RF accelerator. By placing a room-temperature solenoid with a magnetic field of about 1 T downstream of the RFQ, in the medium-energy beam transport (MEBT) section, it is possible to produce vortex protons with an orbital angular momentum (OAM) quantum number $|\ell| \geq 1$, while maintaining acceptable control over emittance growth. This scheme is also applicable to the generation of vortex heavy-ion beams, although optimization of the initial charge states of the ions is more critical in such cases.

At the Institute of Modern Physics, existing room-temperature ion linac facilities can be upgraded to experimentally verify the vortex ion beam generation using the magnetic-immersed stripping method. The primary challenge at present lies in unambiguously confirming the generation of vortex ions. The development of appropriate diagnostic techniques, along with matching beamline designs, constitutes the next major step in this research.

Assessing the feasibility of conducting these studies at accelerator facility is an important step in advancing this research field. The China Advanced Nuclear Physics Research Facility offers versatile conditions for researching the generation, manipulation, detection and application of vortex particles in accelerators. This will lay a solid foundation for future theoretical and experimental studies, particularly in high energy nuclear and particle physics, involving vortex particles.

\section{Summary}
The research facilities in Huizhou, which include the China initiative Accelerator Driven System (CiADS), the High Intensity Accelerator Facility (HIAF), and the China Nuclear User Facility (CNUF), are strategically designed to tackle significant scientific challenges through advanced experimental environments. These facilities aim to enable high-precision investigations across four pivotal areas: rare decays of the $\eta$ meson beyond the Standard Model, muon physics, neutrino phenomena, and cold neutron research. A primary objective is to identify and execute experiments of considerable physical significance, utilizing the unique capabilities of the Huizhou facilities to achieve high-precision measurements while ensuring that detection technologies are adaptable to the diverse demands of various research endeavors.

In $\eta$ meson studies, the proposed Super $\eta$ Factory is expected to revolutionize $\eta$ production, achieving statistical yields approximately two orders of magnitude higher than those currently possible at the Super Tau Charm Factory. This increased production capacity will present unparalleled opportunities for discovering new particles and interactions. Detailed investigations of meson decay patterns will facilitate rigorous examinations of fundamental symmetries, including parity and its violations, potentially illuminating the enduring matter-antimatter asymmetry prevalent in the universe. The facility’s unique capabilities will enable precise tests of the Standard Model and enhance our understanding of non-perturbative QCD, while also shedding light on anomalies such as the Adler-Bell-Jackiw (ABJ) anomaly, which may inform theories on neutrino mass generation under gravitational influences.

In the arena of muon physics, our commitment is to position China at the forefront of high-intensity muon research. By advancing critical technologies and methodically designing experimental approaches, we seek to make significant contributions to the exploration of charged lepton flavor violation (cLFV) and other precision measurements essential for probing physics beyond the Standard Model.

The Huizhou facilities will also offer a comprehensive array of neutrino beams, spanning a wide range of energies to enable in-depth studies of atomic and nuclear structures. This capability will complement ongoing electron-based research at leading institutions, such as Jefferson Lab and the Electron-Ion Collider (EIC). Additionally, we plan to undertake deep inelastic neutrino scattering experiments to obtain precise measurements of fundamental parameters, including the weak mixing angle and the elements of the CKM matrix. The outcomes of these studies are expected to enhance our understanding of novel physics phenomena.

In neutron physics, our focus will center on examining parity violation in nucleons induced by weak interactions, leveraging high-intensity neutron beams generated by the CiADS. Our experiments are designed to tackle the neutron lifetime puzzle and explore baryon number violation, addressing critical issues as outlined by Sakharov.

Overall, the Huizhou Nuclear Science Center is envisaged as a premier research platform that supports both fundamental and applied inquiries, rigorously investigating essential symmetries of the Standard Model while venturing into new physics territories. Through strategic collaboration and a commitment to fostering a diverse pool of top-tier talent, Huizhou aims to establish itself as a leading center for nuclear physics, significantly enhancing China's role in the global landscape of high-precision nuclear physics research and resulting in a substantial contribution to international scientific discourse.

\section*{Acknowledgments}
This project is supported by
National Natural Science Foundation of China under Grant Nos. 12075326; 
Guangdong Basic and Applied Basic Research Foundation under Grant No. 2025A1515010669;
Natural Science Foundation of Guangzhou under Grant No. 2024A04J6243;
Fundamental Research Funds for the Central Universities (23xkjc017) in Sun Yat-sen University;
and Innovation Training Program for bachelor students in Sun Yat-sen University.
The simulation conducted by the MACE working group benefited greatly from the provision of computing resources by the National Supercomputer Center in Guangzhou.

\bibliographystyle{apsrev4-2}
\bibliography{bib}
\end{document}